\documentclass[namedreferences]{SolarPhysics}
\usepackage[optionalrh,solaenum]{spr-sola-addons} 
\usepackage{graphicx}                    
\usepackage{amssymb}                    
\usepackage[usenames]{color}                         
\usepackage{url}                         

\newcommand{\BE}{\begin{equation}}
\newcommand{\EE}{\end{equation}}
\newcommand{\BA}{\begin{eqnarray}}
\newcommand{\EA}{\end{eqnarray}}
 \newcommand{\fig}[1]{Figure~\ref{fig:#1}}
 \newcommand{\figs}[2]{Figures~\ref{fig:#1},\ref{fig:#2}}
 
 \newcommand{\sect}[1]{Section~\ref{sec:#1}}
 \newcommand{\sectss}[2]{Sections~\ref{sec:#1}-\ref{sec:#2}}
 \newcommand{\eq}[1]{Equation~(\ref{eq:#1})}
 \newcommand{\eqs}[2]{Equations~(\ref{eq:#1},\,\ref{eq:#2})}

\newcommand{\uvec}[1]{ \hat{\bf #1} }

\newcommand{\arcsec}{''} 
\newcommand{\degree}{^{0}} 
\newcommand{\FeXII}{Fe {\sc xii}} 
\newcommand{\FeXIIA}{Fe {\sc xii} $195.12$~\AA}  
\newcommand{\FeXV}{Fe {\sc xv}} 
\newcommand{\FeXVA}{Fe {\sc xv} $284.16$~\AA}  
\newcommand{\SiVII}{Si {\sc vii}} 
\newcommand{\SiVIIA}{Si {\sc vii} $275.35$~\AA}  
\newcommand{\kms}{km s$^{-1}$} 
\newcommand{\ur}{\uvec{u}_{\rm r}}
\newcommand{\ut}{\uvec{u}_{\rm \theta}}
\newcommand{\up}{\uvec{u}_{\rm \varphi}}
\newcommand{\Vr}{V_{\rm r}}
\newcommand{\Vt}{V_{\rm \theta}}
\newcommand{\Vp}{V_{\rm \varphi}}
\newcommand{\Vo}{V_{//}}
\newcommand{\Wr}{W_{\rm r}}
\newcommand{\Wp}{W_{\rm \varphi}}

\newcommand{\eg}{\textit{e.g.}}
\newcommand{\ie}{\textit{i.e.}}

\begin{document}
\begin{article}
\begin{opening}
\title{The 3D geometry of active region upflows deduced from their limb-to-limb evolution}

\author{
P.~\surname{D\'emoulin}$^{1}$    \sep
D.~\surname{Baker}$^{2}$    \sep 
C.H.~\surname{Mandrini}$^{3,4}$      \sep      
L.~\surname{van Driel-Gesztelyi}$^{1,2,5}$  
       }

%
\runningauthor{P. D\'emoulin \textit{et al.}}
\runningtitle{Longterm evolution of active region flows}

%
  \institute{$^{1}$ Observatoire de Paris, LESIA, UMR 8109 (CNRS), F-92195 Meudon Principal Cedex, France \\ 
  Corresponding author P. D\'emoulin, email: \url{pascal.demoulin@obspm.fr}\\ 
             $^{2}$ UCL-Mullard Space Science Laboratory, Holmbury St Mary, Dorking, Surrey, RH5 6NT, UK\\
             $^{3}$ Instituto de Astronom\'\i a y F\'\i sica del Espacio (IAFE), CONICET-UBA, Buenos Aires, Argentina\\
             $^{4}$ Facultad de Ciencias Exactas y Naturales (FCEN), UBA, Buenos Aires, Argentina\\
             $^{5}$ Konkoly Observatory, Research Centre for Astronomy and Earth Sciences, Hungarian Academy of Sciences, Budapest, Hungary}
             
\begin{abstract}
We analyse the evolution of coronal plasma upflows from the edges of AR 10978, which has the best limb-to-limb data coverage with \textit{Hinode's EUV Imaging Spectrometer} (EIS).  We find that the observed evolution is largely due to the solar rotation progressively changing the viewpoint of nearly stationary flows. From the systematic changes in the upflow regions as a function of distance from disc centre, we deduce their 3D geometrical properties as inclination and angular spread in three coronal lines (\SiVII , \FeXII , \FeXV ). In agreement with magnetic extrapolations, we find that the flows are thin, fan-like structures rooted in quasi separatrix layers (QSLs). The fans are tilted away from the AR centre. The highest plasma velocities in these three spectral lines have similar magnitudes and their heights increase with temperature. The spatial location and extent of the upflow regions in the \SiVII , \FeXII\ and \FeXV\ lines are different owing to (i) temperature stratification and (ii) line of sight integration of the spectral profiles with significantly different backgrounds.  We conclude that we sample the same flows at different temperatures.  Further, we find that the evolution of line widths during the disc passage is compatible with a broad range of velocities in the flows.  Everything considered, our results are compatible with the AR upflows originating from reconnections along QSLs between over-pressure AR loops and neighboring under-pressure loops. The flows are driven along magnetic field lines by a pressure gradient in a stratified atmosphere. 
We propose that, at any given time, we observe the superposition of flows created by successive reconnections, leading to a broad velocity distribution. 
\end{abstract}
%
\end{opening}

\section{Introduction}\label{sec:intro}

An extensive body of literature exists discussing active region (AR) plasma motions and Doppler shift measurements of chromospheric, transition region (TR), and to a lesser extent, coronal emission lines from the spectrometers aboard rocket flights, \textit{Skylab}, \textit{Solar Maximum Mission} (SMM) \textit{Ultraviolet Spectrometer and Polarimeter} (USVP), and the \textit{Coronal Diagnostic Spectrometer} (CDS) and \textit{Solar Ultraviolet Measurements of Emitted Radiation} (SUMER) on board of the \textit{Solar and Heliospheric Observatory} (SOHO), among others (see e.g., \opencite{Brynildsen98}; \opencite{teriaca99}; \opencite{Marsch04}; and references therein).

\subsection{General Characteristics of AR Upflows}  
Previous measurements of AR plasma flows have been limited by the spectral, spatial and temporal resolution of spectrographs. Since its launch on-board the \textit{Hinode} satellite \cite{Kosugi07} on 23 September 2006, the \textit{EUV Imaging Spectrometer}  (EIS, \opencite{Culhane07}) has produced routine measurements of Doppler shifts and broadening in lines formed at TR and coronal temperatures.  AR upflows are especially important because they are considered to be a possible source of the slow solar wind (SW) \cite{Sakao07,Harra08,DelZanna11,Brooks11}.

   One of the most intriguing EIS results is the detection of high-speed upflows in coronal plasma, $T \sim$ 1~MK, at the edges of ARs \cite{Doschek07,Doschek08,DelZanna08,Harra08,Hara08}.  Each persistent and long-lived upflow is located in a region of low electron density and low radiance over strong magnetic flux concentrations of a single polarity. 
The blueshifts in line profiles range from a few to 50 \kms\ and they are faster in hotter coronal emission lines \cite{DelZanna08}.  A similar dependence of upflow speed magnitude on temperature is also observed in dimming regions which result from a CME (\eg, \opencite{Imada07}; \citeyear{Imada11}).

Recent observations indicate a complex distribution of upflows. For example,
\inlinecite{DelZanna08} and \inlinecite{Tripathi09} identified specific locations within ARs where there is a bifurcation of plasma flows from the TR to the corona.  Doppler velocity measurements in fan loops decrease with increasing temperature in lines formed at 0.6-0.8~MK  (\textit{i.e.} Fe {\sc vii}, Fe {\sc viii}, and Fe {\sc ix}; \opencite{DelZanna09a}; \citeyear{DelZanna09b}).  
In a study of the temperature dependence and morphology of an AR upflow region, \inlinecite{Warren11} showed that the velocity structure is highly complex with upflows observed in emission lines from Fe {\sc xi} to Fe {\sc xv} and downflows in bright fan-like loops at the lower temperature of Si {\sc vii}.  \inlinecite{Ugarte11} and \inlinecite{Young12} found a similar velocity pattern in ARs across the temperature range of 0.15-1.5~MK.  
\inlinecite{Warren11} suggest that the fan loops and the strong upflow regions form two different independent populations.

\subsection{Blue-Wing Asymmetry}
Recent work, based on EIS observations, by \inlinecite{Hara08}, \inlinecite{DePontieu09}, \inlinecite{McIntosh09}, \inlinecite{Peter10}, and \inlinecite{Bryans10} among others, has provided new insight into the nature of emission line profiles of hotter coronal lines, especially EIS's core Fe  {\sc xii} line, Fe {\sc xiv}, and Fe {\sc xv}. \inlinecite{Hara08} first reported that significant deviations from a single-Gaussian profile were found in the blue wings of the Fe {\sc xiv} and Fe  {\sc xv} line profiles for upflows near AR footpoints.  These upflows contain a high-velocity component of more than 100 \kms .  \inlinecite{DePontieu09}, \inlinecite{McIntosh09}, \inlinecite{Tian11a}, \inlinecite{Tian11b}, 
and \inlinecite{Sechler12} have used maps of blue-red (B-R) asymmetry in the hotter Fe emission lines to isolate bright line core upflows, with typical velocities of
10-20 \kms , from the faint upflows in the blue wings that have velocities of 50-100 \kms . 

Of particular interest to \inlinecite{Bryans10} is the correlation between Doppler shifts and line widths reported by \inlinecite{Doschek08} and \inlinecite{Hara08}. \inlinecite{Bryans10} suggest that emission lines showing the largest line widths indicate that the upflows may result from multiple flow sites. They model the upflowing plasma by imposing a double-Gaussian fit to the asymmetric line profiles of Fe {\sc xii} and Fe {\sc xiii}.  In their limb-to-limb study of AR 10978, the primary component median velocities were 0-10 \kms\ for the eastern upflow region and 5-13 \kms\ for the western upflow region. Secondary component median velocities were 90-120 \kms\ on the East and 100-130 \kms\ on the West, with over 200 \kms\ observed.  Significant contribution from the secondary components occurred in isolated regions at the base of the upflows on both sides of the AR and were observed throughout the entire seven day period. 
Even though a double-Gaussian fit of the Fe  {\sc xii} and Fe  {\sc xiii} emission line profiles gives a more accurate fit compared to a single-Gaussian fit, \inlinecite{Bryans10} still found a correlation between velocity and width for the primary component. This suggests that a double-Gaussian fit does not fully describe the upflow regions.  Finally, lower temperature lines including Si {\sc vii} and Fe {\sc viii} were found to have symmetric line profiles that are well modeled by a single Gaussian \cite{Bryans10}. 

Though interpretation of the asymmetries in the blue wing of coronal spectral line profiles is still open to significant debate,  \inlinecite{Peter10} suggests that line profile asymmetry can be, among other posibilites, due to: two or more spatial components, opacity effects, non-Maxwellian velocity distribution of ions, line blends, and asymmetric instrument profile.

\subsection{Paper Road Map}
  In the present work we continue the analysis of AR 10978 by \inlinecite{Bryans10}.  This AR is selected because it has the longest temporal coverage of an AR in EIS observations during its  disc transit in December 2007. Moreover, the full spatial extent of the AR was covered by most of the scans. Such complete temporal and spatial coverage remain exceptional in EIS data though a few other ARs also have been well-covered by EIS (see, \eg\ \citeauthor{DelZanna08} \citeyear{DelZanna08}, \citeyear{DelZanna11}).  
  
Our aim is to use this exceptional EIS data set to constrain the geometry, nature, and physics of large-scale upflows present on both sides of ARs.  
We address key questions such as:  How are these flows oriented? How broad is the angular extent of the flows in both North-South and radial directions? Are upflows observed in different spectral lines related? How are they linked? How dispersed are the flow speeds?

  In \sect{data}, we summarise the main observed flow characteristics of AR 10978, then, we describe the stationary flow model used to separate the apparent evolution due to solar rotation from the intrinsic velocity evolution.
   In \sect{Inclin}, we analyse the Doppler blue-shift (tracing upflows) as derived from a single Gaussian fit to the profiles of three spectral lines.  From their evolution with
solar rotation, we derive the inclination of the flows to the local vertical.   
For the strongest line, \FeXIIA , we perform the same analysis for the highest flows derived from a double Gaussian fit of the spectral profile.  Finally, we compare the different flow inclinations to the results of magnetic extrapolations.
   In \sect{Spread}, we relate the upflows observed in the three spectral lines and we derive the upflow spread both in angular directions and velocity magnitude.
More precisely, the observed evolution of the angular width provides independent information on the upflow inclination to the local vertical as well as the horizontal angular spread. Next, we use forward modelling to constrain the vertical angular spread of the upflows. Finally, the evolution of the line width with solar rotation permits us to infer the physical origin of the line broadening. 
  We end by summarising our results and drawing synthesised conclusions in \sect{Conclusion}.

\begin{figure}  
 \centerline{ \includegraphics[width=1.\textwidth]{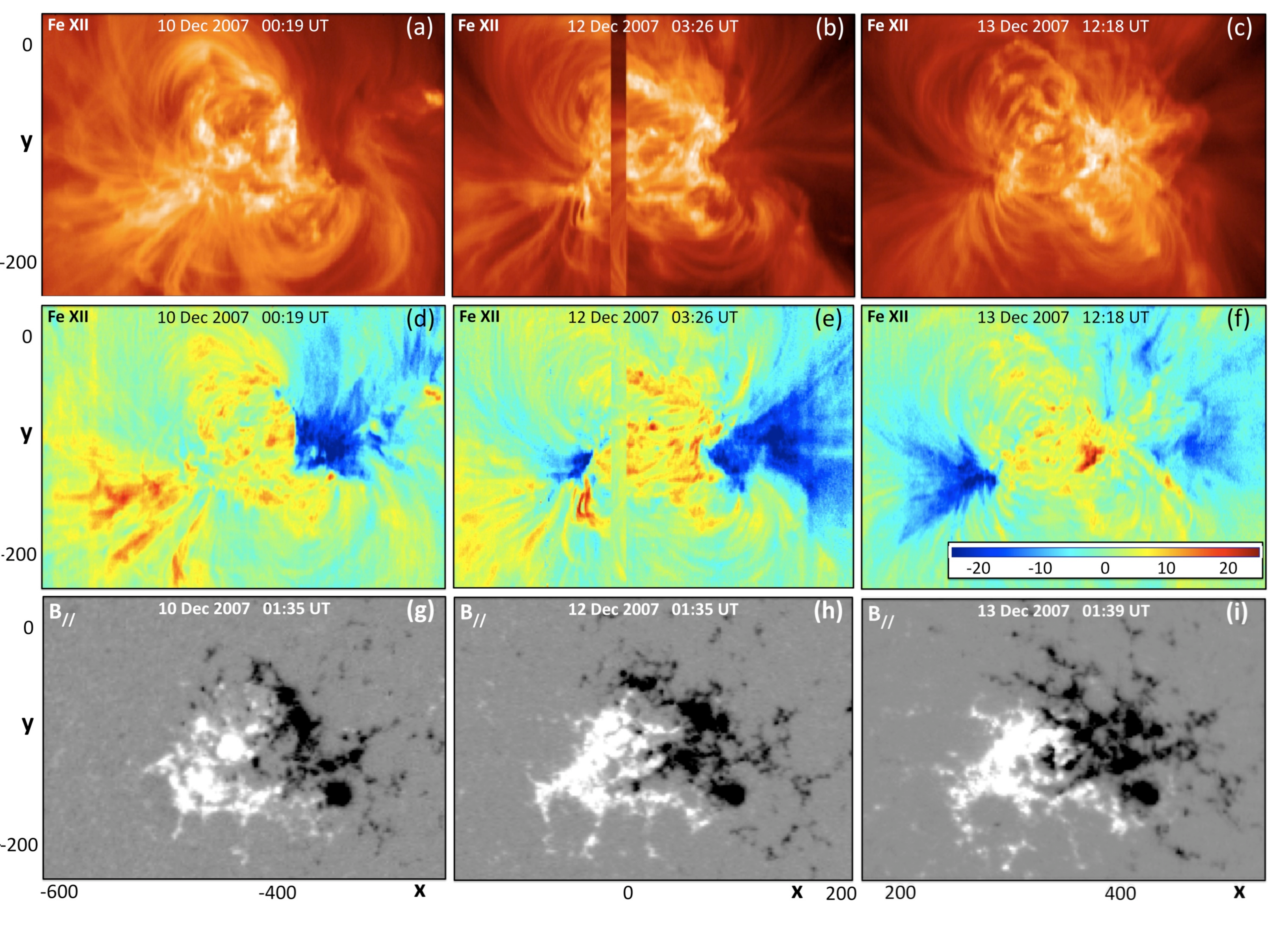} }
     \vspace{-0.03\textwidth}   
\caption{Observed evolution of AR 10978 over three days: 10 Dec. 00:19 UT, 12 Dec. 03:26 UT and 13 Dec. 12:18 UT (selected close to central meridian passage (CMP) and on both sides when AR 10978 was at $X \approx \pm 400$\arcsec).  The top and middle rows show the intensity and the Doppler velocity deduced from a single Gaussian fit of the \FeXIIA\ line (the corresponding full set of data is shown in movie iv\_4panels*).  The bottom row shows the MDI magnetograms closest in time with a 500~G level saturation.
}
 \label{fig:IvB_maps}
\end{figure}  

\begin{table}
\caption{EIS study details.  All rasters are constructed with the 1\arcsec\ slit.}
\begin{tabular}{ccrrcccc}
\hline
Date  & Start & \multicolumn{2}{c}{Location} & Study & Exposure & field of view & Raster \\ 
      & Time&   X      &      Y   &       &Time&X$\times$Y& Time \\ 
(2007)& (UT)&(\arcsec )&(\arcsec )& (No.) &(s) &(\arcsec )& (hr) \\ 
\hline
09 Dec &18:23 & -554 & -175 &233 &10&360 $\times$ 512 &1.0\\
10 Dec &00:19 & -447 & -161 &198 &40&460 $\times$ 384 &5.1\\
11 Dec &00:24 & -178 & -144 &~54 &60&256 $\times$ 256 &4.3\\
11 Dec &10:25 & -150 & -141 &198 &40&460 $\times$ 384 &5.1\\
11 Dec &16:24 &  -65 & -118 &~46 &15&256 $\times$ 256 &1.1\\
12 Dec &03:26 &   -9 & -148 &198 &40&460 $\times$ 384 &2.3\\
12 Dec &11:43 &   91 & -128 &198 &40&460 $\times$ 384 &2.3\\
13 Dec &12:18 &  347 & -134 &198 &40&460 $\times$ 384 &5.1\\
15 Dec &00:13 &  620 & -142 &198 &40&460 $\times$ 384 &5.1\\
15 Dec &18:15 &  737 & -142 &198 &40&460 $\times$ 384 &5.1\\
\hline
\end{tabular}
\label{eis_info}
\end{table}

\section{Data and Analysis Method}\label{sec:data}

\subsection{Observations and Data Reduction} \label{sec:EIS}

{\it Hinode} EIS tracked AR~10978 from 6 to 19 December 2007.  EIS's full complement of slits was utilised to perform both raster and sit-and-stare observations.  These observations provide the most extensive spectroscopic coverage of an AR crossing the solar disc.  Here we concentrate on ten EIS rasters constructed with sufficiently large fields of view to cover both AR magnetic polarities (\fig{IvB_maps}).  Our overriding aim is to obtain velocity and line width information for a wide temperature range of emission lines in order to analyse the evolution of flows from AR~10978.  Raster information is shown in Table~\ref{eis_info} (please see other authors for a more detailed description of the EIS observations for AR 10978 {\emph e.g.} \opencite{Doschek08}; \opencite{Bryans10}; \opencite{Brooks11}).

EIS data were processed using standard Solar Soft EIS routines eis$\_$prep and eis$\_$auto$\_$fit.  Raw data were corrected for dark current, cosmic rays, hot, warm and dusty pixels.  Instrumental effects of slit tilt and orbital variation in the line centroid position due to thermal drift were removed.  In the cases of \SiVIIA\ and \FeXVA\ data (T$ \approx$ 0.6 and 2~MK, respectively; \opencite{Young07a}), calibrated spectra were fitted with a single gaussian function in order to obtain the line centre for each spectral profile.  Reference wavelengths were taken from an average value obtained in a relatively quiescent Sun patch of each raster.  Finally, \SiVIIA\ data were rebinned $4 \times 4$ in order to improve signal over noise.

For the core EIS \FeXII\ emission line at 195.12~\AA ($T \approx 1.4$~MK; \opencite{Young07a}; \opencite{DelZanna08}), calibrated spectra were fitted using both single and double gaussian functions.  The latter analysis, with more free parameters in the fitting function, is only performed with Fe {\sc xii} emission line because of its stronger signal over noise ratio.
The method used for modelling line profiles as the sum of two gaussians with the same width and a linear background is described in \inlinecite{Bryans10}. 

\begin{table}
\caption{Attached movies.  The data are obtained with \SiVIIA\ (rebinned $4 \times 4$), \FeXIIA\ and \FeXVA\ spectral lines. All velocities are derived from a single Gaussian fit of the line profiles 
and they are shown with the same color table within the range $-25<V_{//}<25$ \kms . Only the beginning of the movie name is given.}
\begin{tabular}{ll}
\hline
Name & Content\\ 
\hline
i\_compare*    & intensity comparison of the three spectral lines \\
i\_evolution*  & intensity evolution for each spectral line \\
iv\_4panels*   & Top panels: intensity and velocity of the \FeXII\ line\\
               & Bottom panels: velocities of the \SiVII\ and \FeXV\ lines  \\
iv\_compare*   & comparison of intensity to velocity for each spectral line \\
v\_compare*    & velocity comparison of the three spectral lines  \\
v\_evolution*  & velocity evolution for each spectral line \\
\hline
\end{tabular}
\label{movies_info}
\end{table}

   Extreme care has been taken to coalign the different data sets so that the evolution of the flows, both temporal and wavelength-dependent, can be studied to the required level of detail. The results are shown in several movies included as electronic supplement to this article (see the movie list in Table~\ref{movies_info}). Since coalignment using information in the file headers was insufficient, we first utilised the intensity maps by aligning loops and moss regions (see movies i\_compare* and i\_evolution* for the coalignment between spectral lines and between different times, respectively), then we confirmed the results in the velocity maps (movies iv\_compare*, v\_compare*). In particular, the \FeXII\ line is not on the same detector as the two other lines.  As a consequence, a variable shift in the North-South direction is present for all times (typically $\Delta y \approx -15\arcsec,-20$\arcsec), while no extra shift is needed in the East-West direction, except on 12 Dec 2007 03:26 UT ($\Delta x \approx 27$\arcsec).  Constructing multi-wavelength movies of intensity and velocity maps was especially useful to check the coalignment and to understand the geometry of the flows (movie iv\_compare*). The same color table with the same saturation values ($\pm 25$ \kms ) is used for all velocity maps so that they can be easily compared.  We add guide marks to help in tracking the main flow structures. These marks either encircle the flows (black circles/ellipses), or trace the location of a weak velocity pattern (dashed black lines), or separate flow streams (dashed red/blue lines on the western AR side). 
   
   In order to coalign EIS data with MDI magnetograms, we started with the header information; then, we checked/refined the coalignement by comparing the positions of all the intensity structures (moss, loops) with the magnetic polarities (a moss region is expected above a polarity with intermediate field strength, and a loop is expected to link two opposite magnetic polarities).

\begin{figure}  
\centerline{\includegraphics[width=\textwidth]{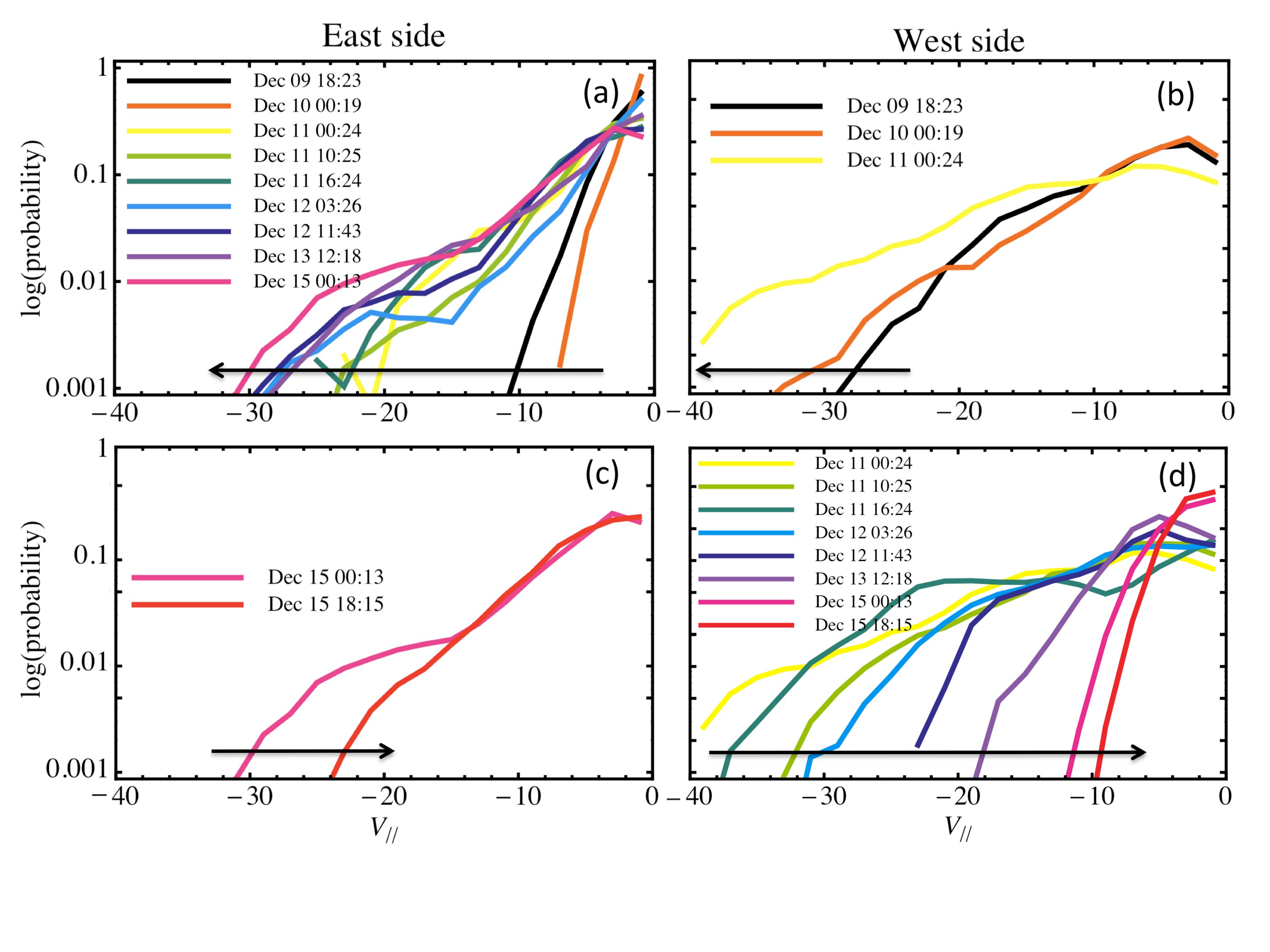}}
     \vspace{-0.08\textwidth}   
\caption{Histograms of the upflow velocities on the eastern and western edges of AR 10978 (they are drawn with continuous lines).  The Doppler velocity is deduced from a single Gaussian fit of the \FeXIIA\ line.  The bin width is 2~\kms\ and the probability of the bins is defined by dividing the counts by their sum.  Due to the large dynamics, a logarithm ordinate is shown.  The distributions of upflows show a monotonic evolution in each panel (as indicated by the arrows) except on the West side for Dec. 11 at 10:25 UT which has slightly lower velocities.
}
 \label{fig:hist_single}
\end{figure}  

\subsection{Overview of flow evolution}\label{sec:Overview}

  As AR 10978 rotated on the solar disc, there is a distinct global evolution of the Doppler velocities, as summarized in movie iv\_4panels*.  On the eastern side the upflow velocities become larger with time, while the reverse evolution is dominant on the western side (\fig{IvB_maps}).  Such evolution progresses with time as shown in the velocity movies in all three spectral lines. 
 
\begin{figure}  
\centerline{\includegraphics[width=0.68\textwidth]{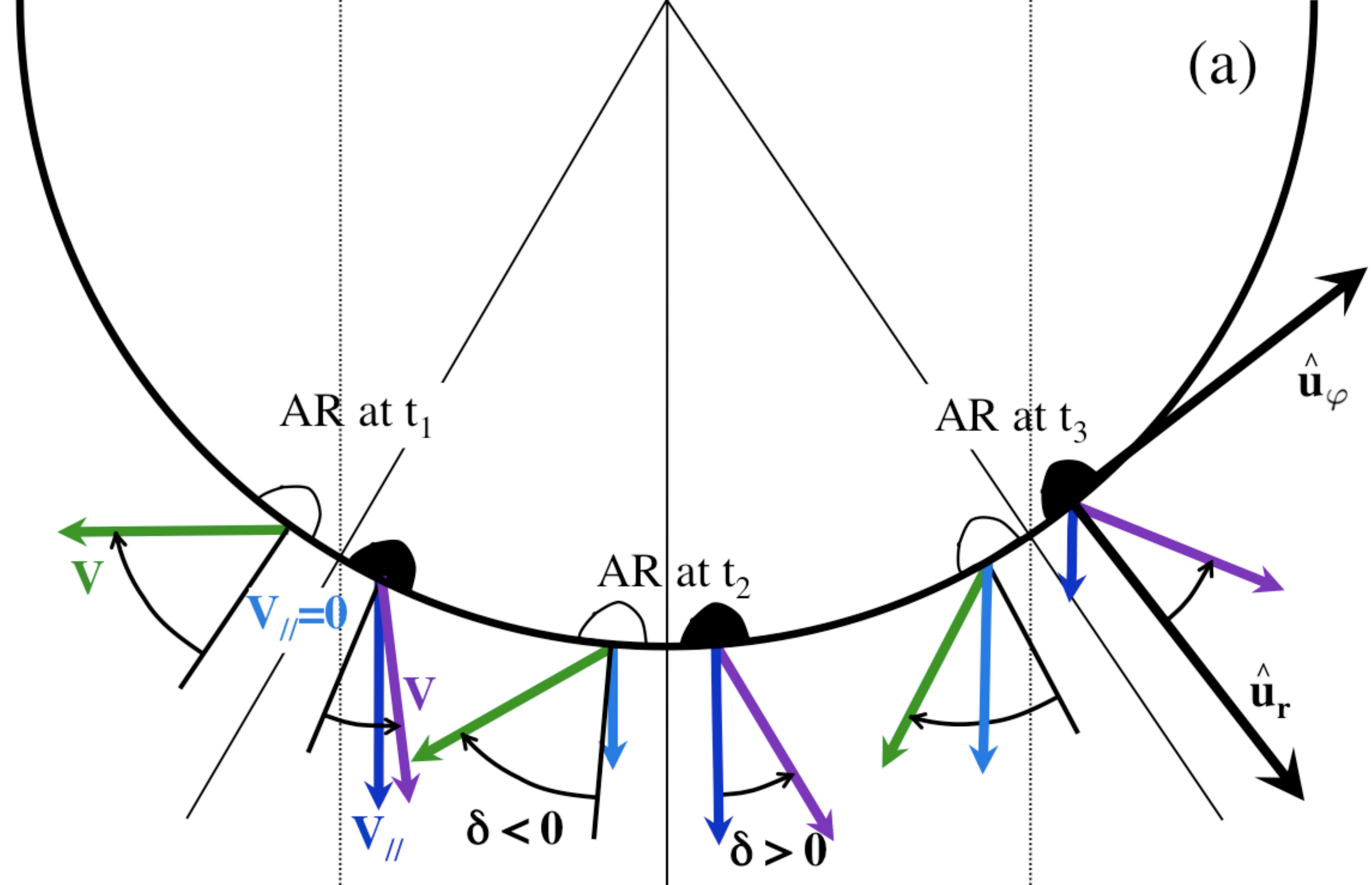} ~~~
            \includegraphics[width=0.29\textwidth]{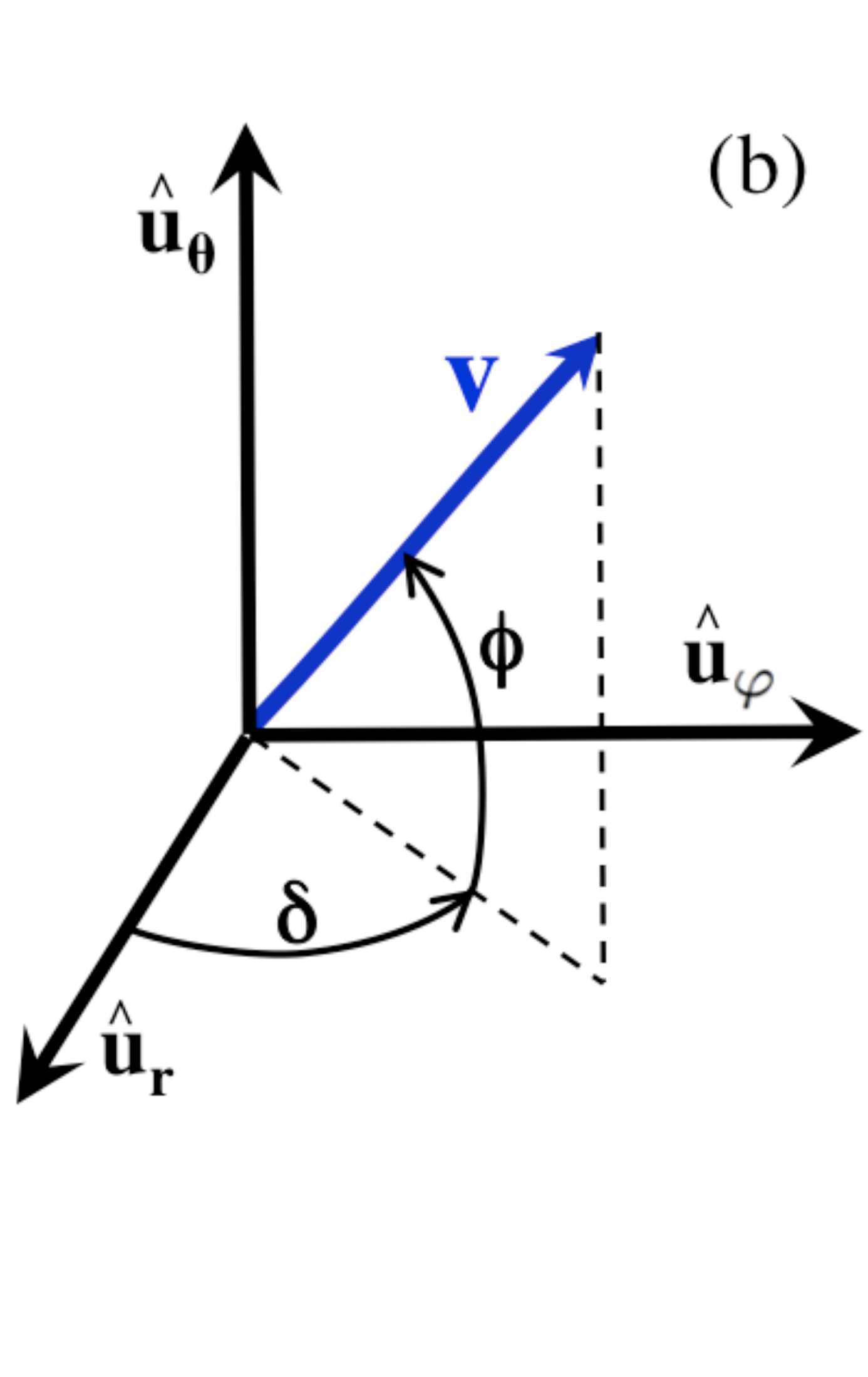} }
\caption{ {\bf (a)} Schema showing the AR at three positions (as viewed from the North Pole). 
The purple and green arrows show the mean upflow in the leading and following AR polarity, respectively.  The inclination of flows to the vertical are based on a modelling result found by a least square fit to the data (Section~\ref{sec:Inclin}).  The medium and dark blue arrows show the observed Doppler velocities ($V_{//}$).  $\delta$ is the East-West (along $\up$) inclination of the flows to the local vertical ($\ur$).  This schema interprets qualitatively the histogram evolution shown in \fig{hist_single} as well as the velocity movies.
           {\bf (b)} Local system of coordinates for the velocity, as defined by \eq{Vlocal}. $\phi$ is the inclination of the flows in the polar direction ($\ut$).
}
 \label{fig:schema}
\end{figure}

  The evolution is quantified by analysing the histograms of the upflows (\fig{hist_single}). The eastern side shows almost continuous increase in the magnitude 
of the upflow velocities up to the beginning of 15 December (\fig{hist_single}a), while the next observation shows a decrease of the strong upflows (\fig{hist_single}c).  The western side shows a similar evolution, but the change from increasing to decreasing upflow magnitude occurs earlier in time since the strongest upflows are observed at the beginning of 11 December (\fig{hist_single}b,d).  This global evolution is a clear signature of a projection effect evolving with the AR position on the disc. AR 10978 crosses the central meridian around noon on 11 December; so, we deduce that the flows on the eastern/western sides are inclined towards the East/West compared to the local vertical (so both are tilting away from the AR).  From the times when the maximum Doppler velocities are reached, compared to the time of central meridian passage (CMP), one deduces that the tilt is larger on the eastern side than on the western one (\fig{schema}).  Their mean inclinations and angular spreads are quantified in \sect{Inclin} and \ref{sec:Spread}, respectively. 

\subsection{Modelling Steady Flows}\label{sec:Steady}
  Since the observed flows are strongly affected by the solar rotation, we derive below equations for a stationary flow to quantify the effect of the evolving line-of-sight (LOS) projection.  The flows observed with large difference of time (hours) are not expected to be produced by the same plasma, but we suppose that flows are continuously driven with the same velocity magnitude and orientation in the local solar frame. 
After fitting this simple model to the observations (Section~\ref{sec:Inclin}), the deviations of the data from this fit will indicate the temporal evolution of the flows.
      
   We describe the position in spherical coordinates with the radial direction, r, the longitude $\varphi$ and the latitude, $\theta$.  Then, the velocity can be written in general as:
  \BE \label{eq:Vdef} 
  \vec{V} = \Vr \,\ur + \Vp \,\up + \Vt \,\ut \, , 
  \EE
where $\ur$, $\up$, $\ut$ are the unit vectors in spherical coordinates.
The position on the solar disc, normalized to the solar radius, is ($X$: East-West, $Y$: South-North):
  \BA 
   X &=& \sin \varphi \cos \theta   \, ,  \label{eq:XY} \\
   Y &=& \sin \theta      \, . \nonumber  
  \EA

In December 2007, the Sun's rotation axis is inclined by only $0.4\degree$ orthogonally to the Sun-Earth line therefore, we neglect this small angle.  The observed velocity component, $\Vo$, is:
  \BA 
  && \Vo = \Vr' \sqrt{1-X^2-Y^2}  - \Vp \frac{X}{\sqrt{1-Y^2}}   \label{eq:V//} \\ 
  \qquad {\rm with} && \Vr'\; = \Vr - \Vt \frac{Y}{\sqrt{1-Y^2}}                     \,. \nonumber
  \EA
As the AR stays at almost the same latitude, \ie\ at constant $Y$, the evolution of $X$ allows us to separate the velocity component in the meridional plane, $\Vr'$, from the component, $\Vp$, along the longitude direction.
Since the AR latitude is small ($Y \approx -0.1$), $\Vt$ has a small contribution in $\Vo$. 
 
 Next, we write the velocity in a local spherical coordinate system (see \fig{schema}b):    
  \BE \label{eq:Vlocal} 
  \vec{V} = V(\cos \delta \cos \phi \,\ur 
            + \sin \delta \cos \phi \,\up 
            + \sin \phi  \,\ut ) \, , 
  \EE
where $\delta$ is the East-West inclination of the velocity to the local vertical direction ($\ur$) and $\phi$ is the velocity inclination from the $\ur,\up$ plane.  The upflows are mainly in the East-West direction (\fig{IvB_maps}, movie v\_evolution*), then $\phi$ is typically small and \eq{Vlocal} simplifies to:
  \BE \label{eq:Vlocal2} 
  \vec{V} \approx V (\cos \delta \,\ur + \sin \delta \,\up ) \, . 
  \EE
Indeed, both low latitude and low $\phi$ values imply that the observed velocity is largely dominated by $\Vr$ and $\Vp$ components, so that we neglect the contribution of $\Vt$ in the following analysis (see \sect{Spread:NS} for further justifications).
 
\subsection{Tracking Steady Flows}\label{sec:Tracking}

   The flows have an unknown 3D geometry which affects their observed spatial extent as the AR crosses the solar disc.  Added to this geometrical effect, we detect only the projection along the LOS of the velocity and an intrinsic temporal evolution of the flows is to be expected.  Moreover, upflows are typically present in low-emissivity regions (movie iv\_compare*), implying that the spectral line profiles can plausibly be contaminated by foreground or background emissions.  All together, these factors suggest that it is not possible to track a given upflow structure as the AR crosses the solar disc.  
   
   To overcome these difficulties we select the $N$ data points of each data set which have the highest upflows separately on each side of the AR.  
By taking the highest upflows, we limit the data to the most reliable flow velocities.  If the flows are stationary and parallel, the $N$ data points with the highest upflows should have a Doppler velocity evolving as described by \eq{V//}.  The number $N$ can be set as a function of $X$ to compensate for projection effects.  
For example, if the selected data points originate from a similar coronal height, one can remove the foreshortening effect on the limb sides.  However, the upflow regions are expected to be elongated upward along field lines.  This extent in height implies an increasing observed extent when the AR is closer to the limb, so opposite to the previous foreshortening effect.   Then, without a precise knowledge of the 3D extent of upflows it is not possible to tell which projection effect is dominant; so, we consider that $N$ is independent of $X$. Indeed, we find that the inclinations derived below are only weakly dependent on the value of $N$ so we expect that, when the same flows are tracked, the dependence $N(X)$ has only a weak effect on the results.

\begin{figure} 
\centerline{\includegraphics[width=\textwidth]{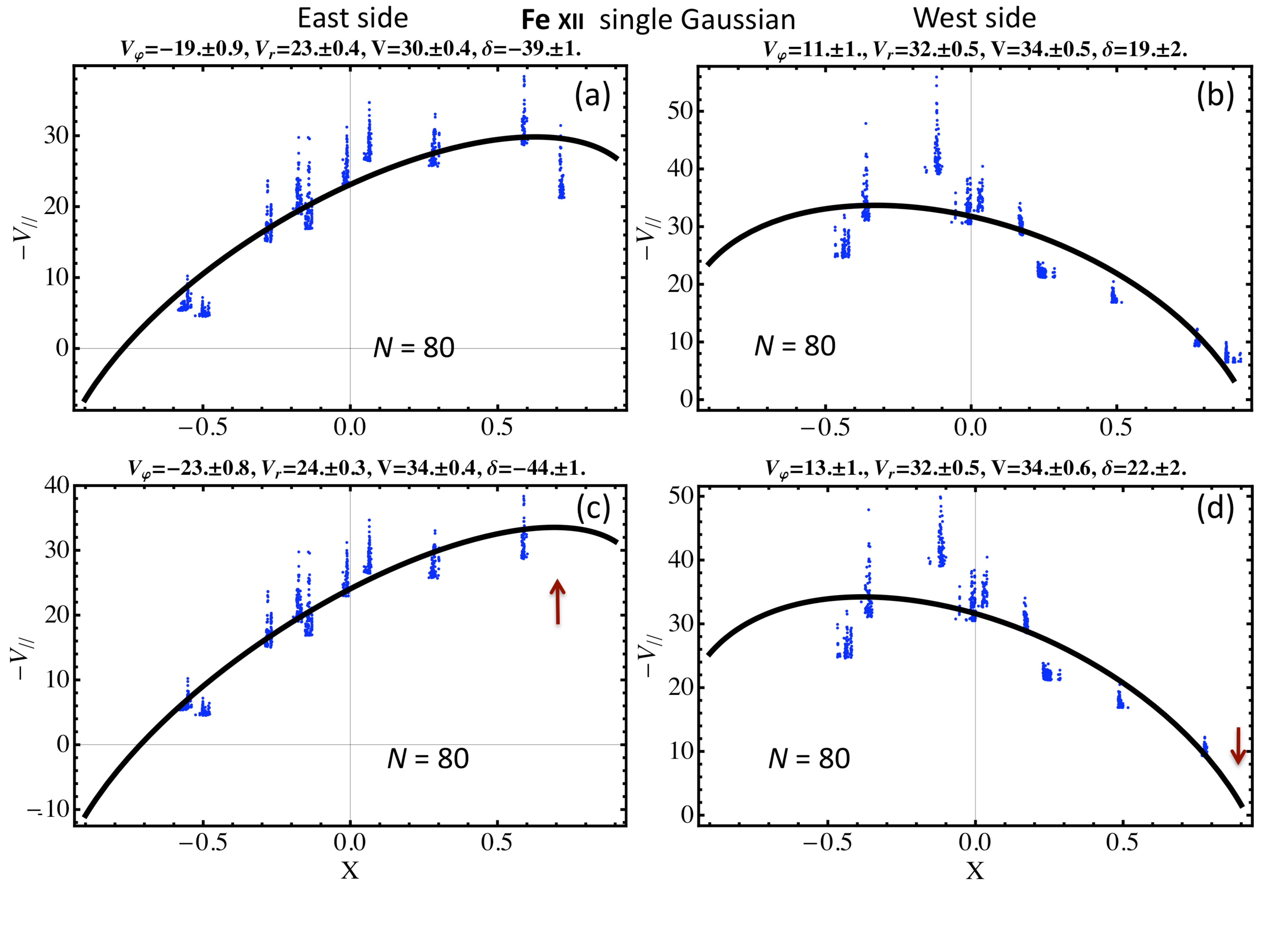}} 
     \vspace{-0.03\textwidth}   
\caption{Dependence of the LOS upflow velocity, $\Vo$, on the $X$ position (normalized to the solar radius) on the eastern (left panels) and western (right panels) sides of the AR.  
$\Vo$ is computed from the spectral shift of single Gaussian fit to the \FeXII\ line.  The 80 highest $-\Vo$ values are selected in each data set (blue points) and the black line is the least square fit of \eq{V//} to these data points.  The results of the fit are written on the top of each panel. In each lower panel the data set with the largest $|X|$ is removed (brown arrow). 
}
 \label{fig:fit_FeXII_single}
\end{figure}  

\section{Mean Flow Inclination to the Local Vertical}\label{sec:Inclin}

\subsection{Inclination Deduced from the \FeXIIA\ line} \label{sec:Inclin:FeXII}

\fig{IvB_maps} and movie v\_evolution* show a fan geometry for the Doppler flows estimated by performing a single Gaussian fit to the \FeXII\ line.  The observed fan geometry and the velocity magnitude, $\Vo$, are both evolving with the AR position on the Sun.  We show below that the evolution of the viewing angle (inclination of the LOS to the local vertical) is dominantly at the origin of the temporal evolution of both of these quantities  (the fan geometry is studied  in \sect{Spread:Spatial}). 

\fig{fit_FeXII_single}a shows that the flows on the eastern side can indeed be described well by \eq{V//}, so they are compatible with nearly stationary flows.   The  backward-leaning (eastward) inclination to the local vertical is significant ($\delta \approx -39\degree$), while the western side is more vertical ($\delta \approx 19\degree$, westward inclination, \fig{fit_FeXII_single}b).  The evolution of the western flows show a significant deviation from \eq{V//}, especially before the central meridian passage.  This corresponds to the flows splitting into two strong upflow structures (see the movie v\_evolution*). Thus, for the western flows, real temporal AR evolution and projection effects have comparable magnitudes, while the evolution of the eastern flows is mostly due to projection effects.  

At least three flow streams can be singled out on the western side.  They are separated in the velocity movies by dashed red lines.  The South-East flow is located in some AR coronal loops (movie iv\_compare*).  This stream has the lowest spatial extent and velocity of the three streams.   The two other streams are similar, only slightly shifted in latitude and in orientation.  Later, these two streams merge (movie v\_evolution*).   We performed a separate analysis for the more East-West oriented of these two streams and we found the results to be comparable to those stated above and shown in \fig{fit_FeXII_single}b (within the quoted error bars).   The South-East flow appears to have only a small contribution, and both of the other streams have a similar temporal evolution.

The uncertainties of $\Vp$ and $\Vr$, quoted on the top of panels in \fig{fit_FeXII_single}, are the errors derived from the least square fit with a 99\% confidence level.  These errors are propagated into $V$ and $\delta$, first by computing the partial derivatives of $V$ and $\delta$ as a function of $\Vp$ and $\Vr$ and second by summing the variances.  

There are other sources of uncertainties.  The number of points retained affects more the eastern than the western mean $\delta$ value; for $N$ in the interval $[20,320]$ the inclination is in $[-42\degree,-38\degree]$, reaching $-47\degree $ for $N=1000$, while on the western side the inclination is stable since it remains within $[19\degree,20\degree]$ for $N$ in the interval $[20,1000]$.  The main effect of $N$ is to decrease $V$ for larger values of $N$, as expected due to the flow selection. $V$ decreases from 37 to 26 \kms\ and from 33 to 22 \kms\ on the eastern and western sides respectively, when $N$ increases from 20 to 1000.  The flow speed is then comparable on both sides. 
   
There are also possible systematic biases.   As expected, the AR latitude has a small effect; doubling the AR latitude ($Y=-200$\arcsec), leads to only a modification of $\delta$ by $1\degree$ on both AR sides.  Next, we have normalized $X,Y$ by the solar radius, supposing that the highest flows originate from low down in the corona.  Redoing the computation with a height of 100\arcsec~above the photosphere has a small effect on $V$ ($\leq 1$ \kms) and increases the inclination of the flows with respect to the vertical (by $3\degree$ on the East side and by $7 \degree$ on the West side). 

Another possible systematic bias is that data taken too close to the limb are distorted by foreshortening and/or overlapping with other structures.  Indeed, a deviation from the fit is seen for the data set with the largest $X$ value (\fig{fit_FeXII_single}a,b).  Another least square fit is performed without this westernmost data set (\fig{fit_FeXII_single}c,d).  On both AR sides, this implies a flow more inclined to the local vertical, by $5\degree$ on the East side and by $3 \degree$ on the West side.

Finally, a significant temporal evolution of the flows is another source of bias. 
This evolution is stronger on the western side especially before the AR crosses the central meridian.  We remove also the data set showing the largest deviation to the fit (first and third data sets in Table~\ref{eis_info}).  The quality of the least square fit to the data becomes comparable to that on the other side (not shown) and the flow is more inclined to the local vertical by $\approx 3\degree$, so again we see no more than a small effect.
Finally, both the limb and temporal effects have negligible influence on the velocity magnitude $V$ (within the error bars).

Within the above uncertainties and biases, we conclude that both AR sides have comparable mean flow speed, $\approx 34$ \kms, with upward directions bent away from the AR, and the eastern side being significantly more inclined to the local vertical than the western side (see Table~\ref{inclinations} for a summary of the main results).

\begin{table}
\caption{Summary of the inclination, $\delta$, found in the eastern and western upflows.
$N$ is the number of pixels retained with the highest upflows. $f=1$ is for original data and $f=16$ for rebinned data. All velocities are obtained with a single Gaussian fit of the line profile except in the most right column where a double Gaussian fit was used (for $I_{\rm secondary}/I_{\rm primary} \geq 0.1$), and the results of the most blue-shifted component are shown.}
\begin{tabular}{ccccccc}
\hline
  $N$  &  \SiVII        &  \SiVII        &  \FeXII   &  \FeXV    & \FeXV      & \FeXII\  \\ 
       & original       & rebinned       & original  & original  & rebinned   & secondary  \\
\hline
\multicolumn{7}{c}{\bf East side}\\
$ 80/f$&    --          &    --          &$-39\pm 1 $&$-55\pm 1 $&$ -56\pm 4 $&$ -22 \pm 3^{~b}$ \\
$160/f$&    --          &    --          &$-40\pm 1 $&$-55\pm 1 $&$ -56\pm 3 $&  --     \\
\multicolumn{7}{c}{\bf East side with the westernmost data removed}\\
$ 80/f$&$-41\pm 3^{~a} $&$-44\pm 10^{~a}$&$-44\pm 1 $&$-56\pm 1 $&$ -59\pm 4 $&$ -25 \pm 3^{~b}$ \\
$160/f$&$-37\pm 4^{~a} $&$-44\pm 10^{~a}$&$-45\pm 1 $&$-57\pm 1 $&$ -59\pm 3 $&$ -26 \pm 3^{~b}$ \\
   &&&&&&\\
\multicolumn{7}{c}{\bf West side}\\
$ 80/f$&    --          &$ 14\pm 8      $&$ 19\pm 2 $&$ 17\pm 5 $&$ 16\pm 1 $&   --    \\
$160/f$&    --          &$ 17\pm 5      $&$ 19\pm 1 $&$ 15\pm 4 $&$ 15\pm 1 $&   --    \\
\multicolumn{7}{c}{\bf West side with the westernmost data removed}\\
$ 80/f$&    --          &$ 23\pm 7      $&$ 22\pm 2 $&$ 21\pm 6 $&$ 20\pm 2 $&$ 19\pm 3^{~c}$ \\
$160/f$&    --          &$ 25\pm 5      $&$ 22\pm 1 $&$ 18\pm 5 $&$ 19\pm 1 $&$ 18\pm 2^{~c}$ \\
\hline
\multicolumn{7}{l}{$^{a}$ The three western data sets are removed (upflows 
hidden by neighbor loops).}\\
\multicolumn{7}{l}{$^{b}$ The two eastern data sets are removed (no data points).}\\
\multicolumn{7}{l}{$^{c}$ The two western data sets are removed (no data points).}\\
\end{tabular}
\label{inclinations}
\end{table}

 \begin{figure} 
\centerline{\includegraphics[width=\textwidth]{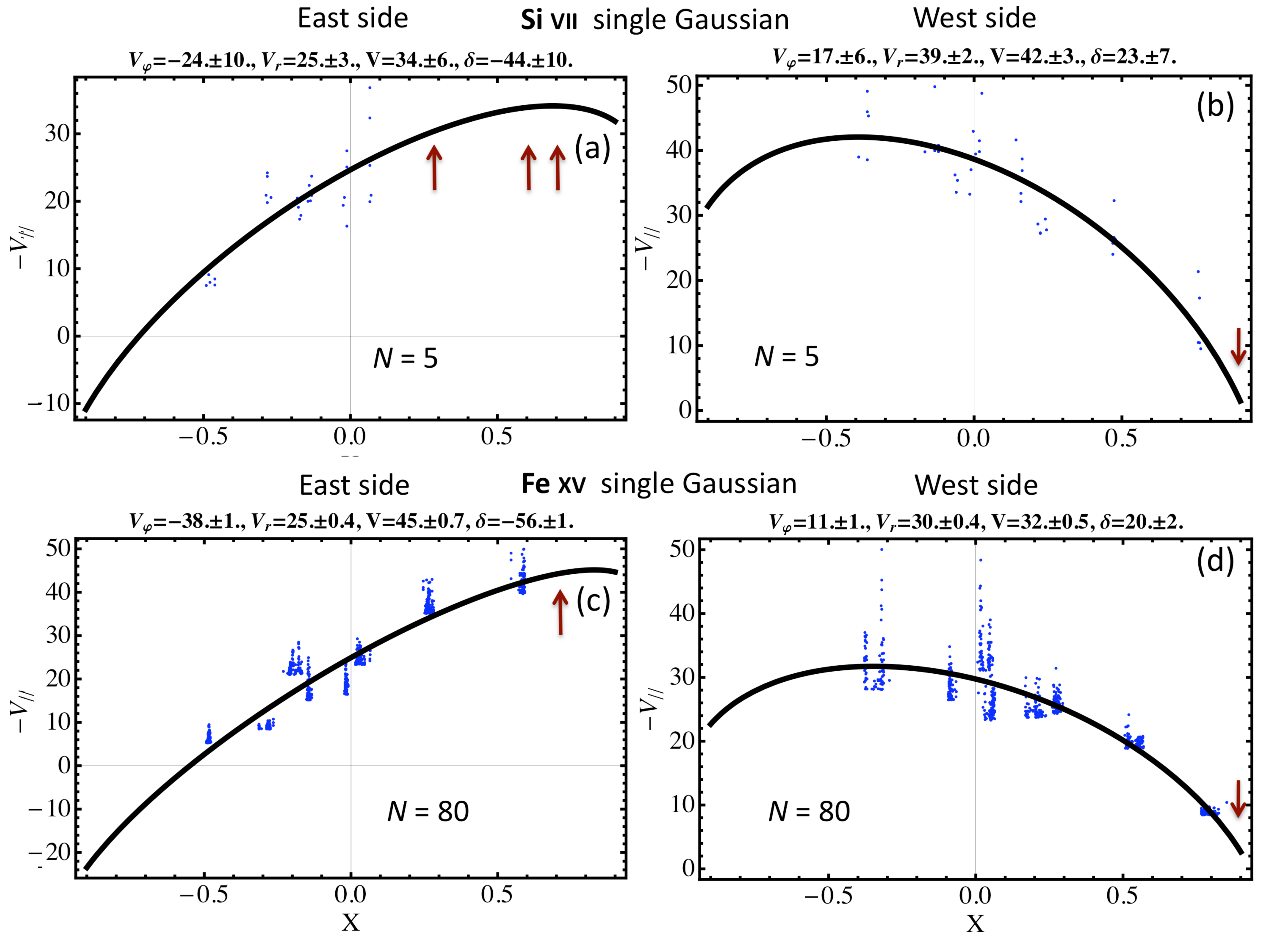}} 
\caption{Dependence of the LOS velocity, $\Vo$, as a function of the $X$ position (normalized to the solar radius) on the East and West edges of the AR, as in \fig{fit_FeXII_single}, but for \SiVII\ (top panels) and for \FeXV\ (bottom panels) with a single Gaussian fit to the spectral profile.  The brown arrows indicate the data set(s) removed from the least square fit of \eq{Vlocal} to the data.
}
 \label{fig:fit_SiVII_FeXV}
\end{figure}  

\subsection{Inclination Deduced from \SiVIIA\ line} \label{sec:Inclin:SiVII}

The \SiVIIA\ line provides information on the flows at a lower temperature (around 0.6~MK) than that of  \FeXIIA\ (around 1.4~MK).  The flows observed in \SiVII\ are less extended and closer to the AR centre than those in \FeXII\ (movie v\_compare*). They have similar temporal evolution, both globally (effect of solar rotation) and locally (presence of three streams on the West side, movie v\_evolution*). 

The \SiVIIA\ line is much weaker than the \FeXIIA\ line so that the spectral profiles are noiser.
Indeed, on the West side no meaningful results can be obtained on a large fraction of the upflows
(the fit of the line profile with a Gaussian function fails).
Then, we improve the ratio of the signal over noise by rebinning the data before fitting the spectral line profile with a Gaussian.   After some experimenting, we select a $4 \times 4$ rebinning (see movies i\_compare* and v\_compare* to compare to the unrebinned data in \FeXII\ and \FeXV ).  On the East side, the signal over noise is strong enough, so rebinning is not needed.  For homogeneity we show the results on both sides with rebinned data in \fig{fit_SiVII_FeXV}a,b.
Results on the East side are comparable with unrebinned data, within the error bars of the fits which are 2 to 3 times lower than with rebinned data (depending on $N$ value, Table~\ref{inclinations}). 
   
Since the upflows observed at different wavelengths are related (see \sect{Spread:Spatial}), the number of points, $N$, needs to be coherently selected to sample comparable flows.  The observations in \SiVII\ are expected to come from regions with stronger magnetic field strength (because lower formation temperature is expected at lower height), so within more concentrated parts of  flux tubes.  However, the emitting extent along flux tubes is not known in such a dynamic situation.  With such knowledge limitation, we select the conservative assumption of $N$ being independent of the observing spectral line.  This turns out to be only a weak limitation as the results in all spectral lines are only weakly dependent on $N$ (as described above for \FeXII , see \sect{Inclin:FeXII}). Finally, to keep the same flows for the analysis, we divide $N$ by 16 for rebinned data. 

As for \FeXII\ (see \fig{fit_FeXII_single}) the westernmost data is slightly away from the least square fit.  Moreover, on the eastern side, after the AR crossed the central meridian, the upflows progressively shrink to a small and elongated corridor (see movie v\_evolution*). This shrinkage to a thin region is consistent with a small spread in inclination of the flows (see \sect{Spread:Vertical}).  However, this implies that the upflows present in this thin volume could be easily mixed up (integrated) with nearby background AR flows, which are dominantly downward flows.  This implies a false weaker upflow signal, as obtained for the last three data sets on the East side.   Therefore, we do not include them in the fit shown in \fig{fit_SiVII_FeXV}a.
   
The inclinations of the flows, $\delta$, found with \SiVII\ are comparable, within the error bars,
with the ones found with \FeXII\ (Table~\ref{inclinations}).  The effect of removing the westernmost data set and/or changing $N$ is also comparable in both spectral lines. The magnitude of the velocity, $V$, is also comparable (even slightly higher in \SiVII\ on the western side, see \sect{Spread:Spatial}).


\subsection{Inclination Deduced from \FeXVA\ line} \label{sec:Inclin:FeXV}

The \FeXV\ line provides information on the flows at a temperature (around 2~MK) higher than that of \FeXIIA\ (around 1.4~MK).  The flows observed in \FeXV\ are more extended and located farther  away from the AR centre than with \FeXII\ (movie v\_compare*).  Otherwise, they have similar spatial shapes and temporal evolution (movie v\_evolution*). 

The \FeXVA\ line is weaker than the \FeXIIA\ line so that spectral profiles are noiser. Unlike for the \SiVII\ line, however, rebinning is not necessary. Indeed, the results obtained with the original and rebinned ($4 \times 4$) data are very similar (within the error bars, Table~\ref{inclinations}). 

The effect of the number of points selected, $N$, as well as the deviation of the westernmost data set from the fit are comparable to the results obtained with \FeXII .  The inclination of the flows, $\delta$, and the velocity magnitude, $V$, are also comparable.  Again, the flows are significantly more inclined to the local vertical for the eastern than the western flows, and this tendency is even enhanced for \FeXV\ (\fig{fit_SiVII_FeXV}c,d, and Table~\ref{inclinations}).

 \begin{figure} 
\centerline{\includegraphics[width=\textwidth]{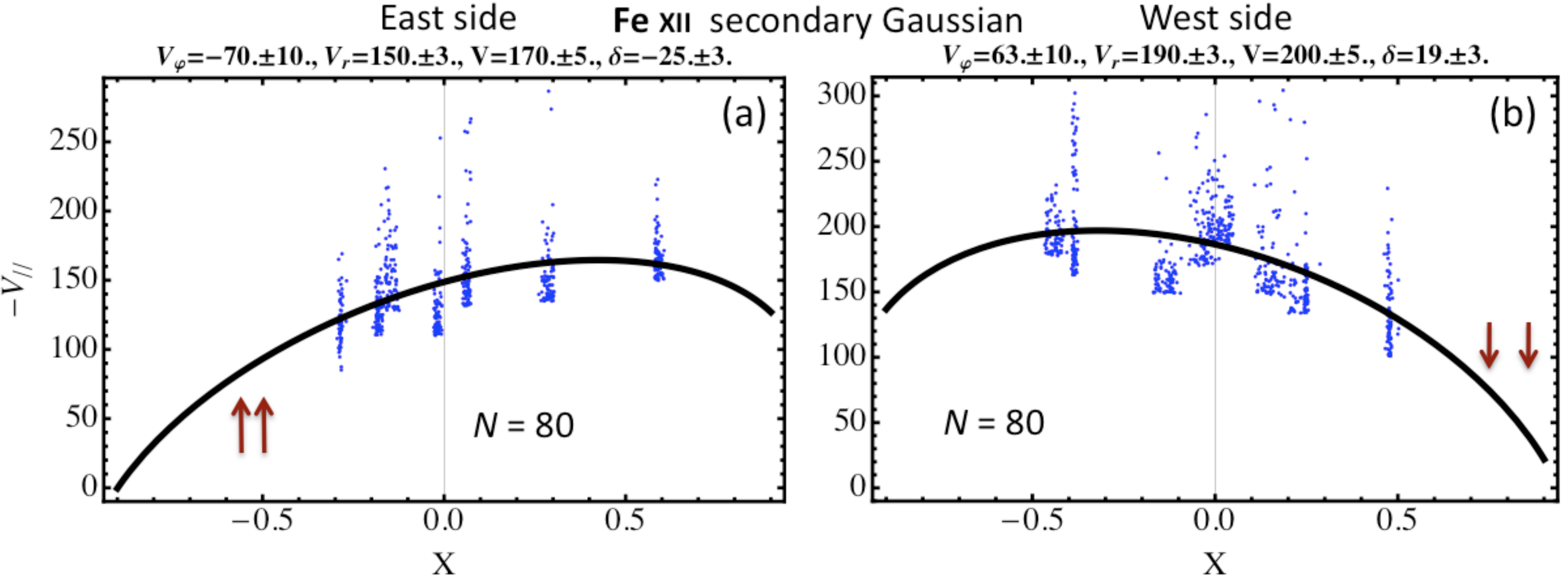}} 
\caption{Dependence of the LOS velocity, $\Vo$, as a function of the $X$ position (normalized to the solar radius) on the East and West edges of the AR.  $\Vo$ is deduced from the spectral Doppler shift of the secondary component (from a fit of two Gaussian functions to the \FeXII\ line).   The brown arrows indicate the data sets removed from the least square fit of \eq{Vlocal} to the data. }
 \label{fig:fit_FeXII_secondary}
\end{figure}  

\subsection{Double-Gaussian Fitting: Inclination of the Faster Flows} \label{sec:Inclin:Fast}

In the upflow regions, the \FeXIIA\ line has an extended blue wing, which is typically interpreted as being due to the presence of spectrally unresolved high-velocity flows.  The fit of the \FeXII\ line with two Gaussian functions is an attempt to isolate the faster flows by separating a secondary component from a primary one.  However, the algorithm finds a secondary component only when the flows have a blue shift typically larger than the half width of the spectral line.  This selects the upflows which are typically larger than 100 \kms\ (see Figures 2 and 7 of Bryans et al. 2010).   Such flows are typically limited to a small portion of the regions where the largest upflows are detected with a single Gaussian fit.

The above selection effect is severe because no secondary component is found for two data sets on each side (the westernmost ones for the western side and the symmetric ones for the eastern side). They are the ones with the lowest detected velocity with a single Gaussian fit (\fig{fit_FeXII_single}). This limits the amplitude of $\Vo$ variation, \textit{i.e.} it implies weaker constraints on the flow inclination.   Furthermore, the selection effect is so severe that the mean velocity of each data set is nearly constant since the detected flows have a velocity just above the threshold imposed by the spectral line width.  Then, the effect of solar rotation on the LOS projection is almost washed out from the corresponding velocity maps as well from a simple analysis of the data.  This is true for any selection criteria adopted for the intensity ratio of the secondary over the primary ($I_{\rm secondary}/I_{\rm primary}$).      
 
Since it has almost stationary flows (\fig{fit_FeXII_single}), the East side of the AR is selected to find out if it is possible to extract some information from the velocity of the secondary component.  We follow the same strategy as previously discussed by selecting the $N$ highest blue shifts in each data set.   Since the number of pixels having a detected secondary component is very limited when the AR is away from the central meridian, a compromise should be reached between analysing a low $N$ value or a low number of data sets.   We present results with $N \leq 80$ and seven data sets.

On the East side, for $I_{\rm secondary}/I_{\rm primary} \geq 0.05$, we found $\delta$ in the interval $[-22\pm 4,-15\pm 3]$.   For $I_{\rm secondary}/I_{\rm primary} \geq 0.1$, $\delta$ is in the interval $[-25\pm 3,-16\pm 5]$, the case with $N=80$ being shown in \fig{fit_FeXII_secondary}a.  This indicates a more vertical flow direction than the one found with a single Gaussian fit (\fig{fit_FeXII_single}, Table~\ref{inclinations}), but a large dispersion of the data points could partly mask the effect of solar rotation.  
  
We proceed the same way for the flows on the West side of the AR. No meaningful results are obtained for $I_{\rm secondary}/I_{\rm primary}\geq 0.05$ since $\delta$ is found fluctuating around $0\degree$. Indeed, the West side has also an additional difficulty, a significant evolution of the flows, as found in \sect{Inclin:FeXII}.  However, the results for $I_{\rm secondary}/I_{\rm primary} \geq 0.1$ are more consistent with those obtained with the single Gaussian fit (\fig{fit_FeXII_secondary}b, Table~\ref{inclinations}).   

We conclude that the velocities of the secondary component need to be severely filtered in order to provide meaningful results.  This implies a low number of useful pixels.   Within these limits, we find that the faster flows have an inclination consistent with the flows derived from a single Gaussian fit (which have a much larger spatial extent). 

The \FeXIIA\ line is emitted by coronal plasma with a temperature in the range $[1.25, 1.5]$~MK (\eg\ \opencite{DelZanna08}; \opencite{Young12}). Then, the sound speed is in the range $[170,200]$ \kms. For $V$ in the range $[200,240]$ \kms, as obtained for $N=80$, the Mach number is in the range $[1.,1.4]$. For the extreme observed values (few points), $V\approx 300$ \kms, the Mach number reaches values in the range $[1.5,1.7]$.  The above ranges are compatible with the simulation results of \inlinecite{Bradshaw11}, see e.g. their Figures 4 and 7, so the observed flows are compatible with a pressure gradient driving supersonic flows.

\begin{figure} 
\centerline{\includegraphics[width=\textwidth]{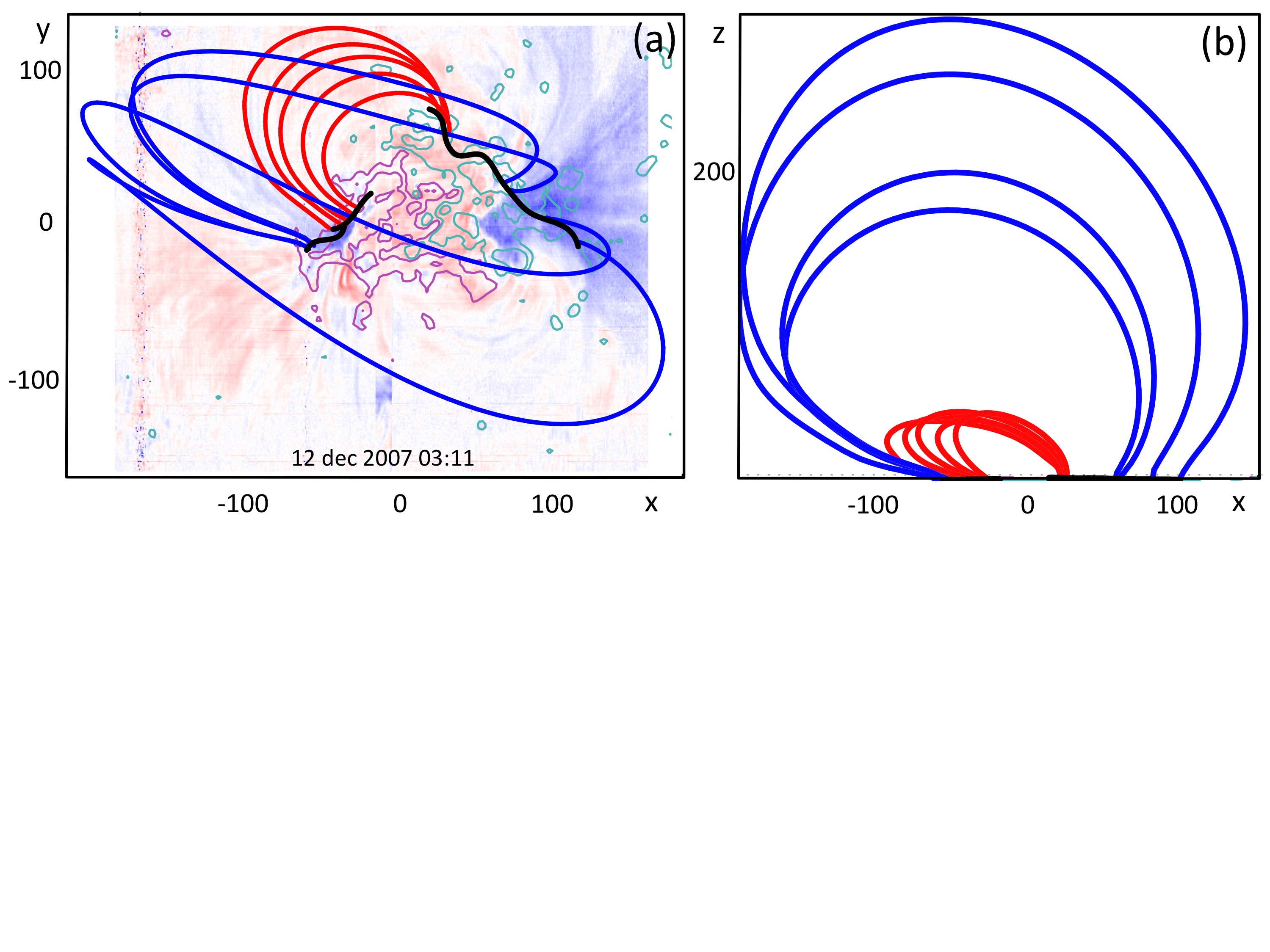}} 
     \vspace{-0.33\textwidth}   
\caption{{\bf (a)} \FeXII\ velocity map overlaid with $\pm 50, 100$~G contours of the longitudinal magnetic field (light blue: negative, pink: positive) and the photospheric trace of QSLs (thick black lines). Field lines are starting from the external side of QSLs at the upflow locations; red field lines have been drawn starting from the East side of the eastern QSL, while blue field lines have their starting integration footpoint on the West side of the western QSL.
{\bf (b)} View from the South pole of the same field lines. 
 }
 \label{fig:extrapol}
\end{figure}  

\subsection{Inclination Deduced from Magnetic Extrapolation} \label{sec:Inclin:Extrapol}

  In order to understand the origin of the emission described in the previous sections and its relation with the three-dimensional AR magnetic structure, we have extrapolated the observed photospheric longitudinal field to the corona using a force-free field configuration ($\vec{J} \times \vec{B} = 0, \vec{\nabla} \times \vec{B} = \alpha B$) and the linear (or constant $\alpha$) force-free field assumption as described in \inlinecite{Demoulin97}.  $\alpha$ is determined in order to minimize the difference between the observed coronal loops and the computed field lines in the AR.
This AR has a magnetic configuration close to potential, so $\alpha$ is small, $\approx -3 \times 10^{-3}$~Mm$^{-1}$.

Next, we compute the quasi-separatrix layers (QSLs), which are thin volumes where the magnetic field connectivity change drastically \cite{Demoulin96a,Masson12}. The upflows are found to be rooted in the photospheric trace of the QSLs and they are mostly following the field lines present on the external side of the QSLs (the sides away from the AR centre, \fig{extrapol}), in agreement with previous studies \cite{Baker09a,vanDriel12}. 

As found previously in \sectss{Inclin:FeXII}{Inclin:FeXV}, the upflows on the western side have an intrinsic evolution (central row of \fig{IvB_maps}).  This is related to the emergence of new bipoles at both sides of the leading polarity (bottom row of \fig{IvB_maps}).  This magnetic field evolution will be studied in a forthcoming paper (Mandrini {\it et al.}, in preparation).
This makes the connectivity of the western polarity more complex than along the eastern polarity. In particular, as can be seen in \fig{extrapol}, the field lines drawn connecting the leading to the following AR polarities do not fully follow the upflows.  A part of these upflows are on field lines connecting the AR leading polarity with a part of the eastern network outside the AR (not shown).  So the upflows involve a more complex set of magnetic connectivities than shown in \fig{extrapol}. 
 The detailed analysis of this relationship will be the subject of a separate paper (Mandrini {\it et al.}, in preparation).  

From 9 to 13 December, when the AR is close to CMP, we perform a magnetic extrapolation on each day.  Then, we compute a set of field lines at the external borders of the QSLs and deduce their East-West inclination, $\delta_{\rm e}$, to the local vertical (\fig{extrapol}).  Averaging the results, on the East side we find $\delta_{\rm e} = -57\degree \pm 8$ which is fully compatible with the inclination deduced from the evolution of the outflows taking into account the error bars (compare to  Table~\ref{inclinations}). On the West side, we find $\delta_{\rm e} = 33\degree \pm 7$, which is between $8\degree$ and $14\degree$ more inclined than the upflows, while the flows are expected to be along field lines (due to frozen-in conditions after reconnection). 
Presently we do not know the origin of this difference (\eg , it could be due to either some limitations of the flow measurements or of the magnetic extrapolations, or both). 

\begin{figure}  
\centerline{\includegraphics[width=\textwidth]{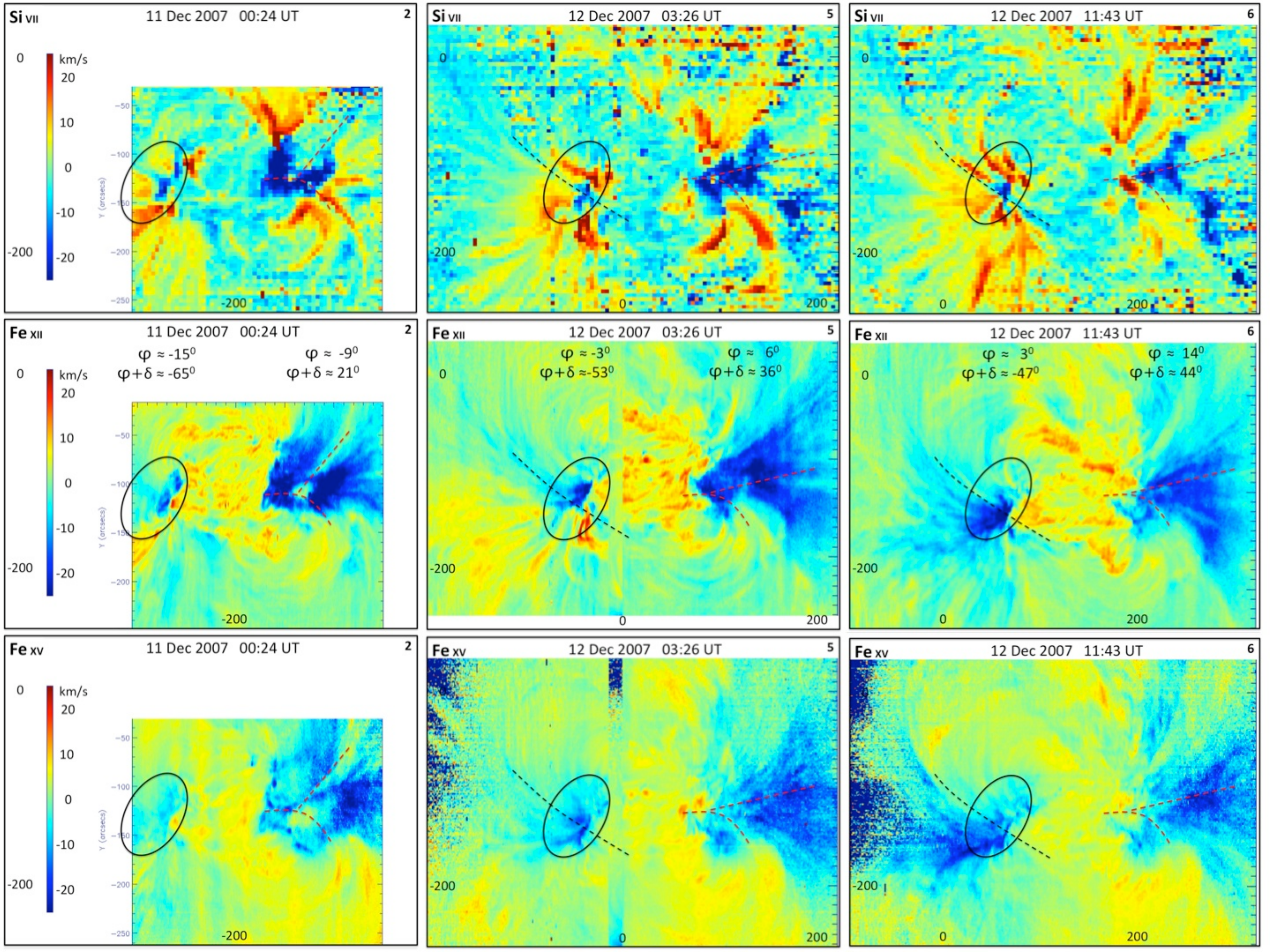}} 
\caption{Comparison of the Doppler velocity deduced from a single Gaussian fit of the \SiVIIA, \FeXIIA\ and \FeXVA\ spectral lines near the central meridian (central column) and at nearly symmetric $X$ positions on both sides (lateral columns).  The same color table with the same saturation values is used for the three spectral lines. In the second row, $\delta$ is the estimated mean inclination to the local vertical, $\varphi$ is the longitude and $\varphi+\delta$ is the mean angle between the flows and the line of sight.  The ellipse and lines mark characteristic features (see \sect{EIS}).  In particular, while the observed flows in \SiVII\ are globally red-shifted,
the blue-shifted counterpart of the flows observed in \FeXII\ and \FeXV\ are marked by an ellipse on the eastern side as they are localized (while they spread out on the western side). The corresponding full set of data is shown in movies v\_compare* and v\_evolution*.
  }
 \label{fig:evol_Si_Fe}
\end{figure}  

\begin{figure}  
\centerline{\includegraphics[width=\textwidth]{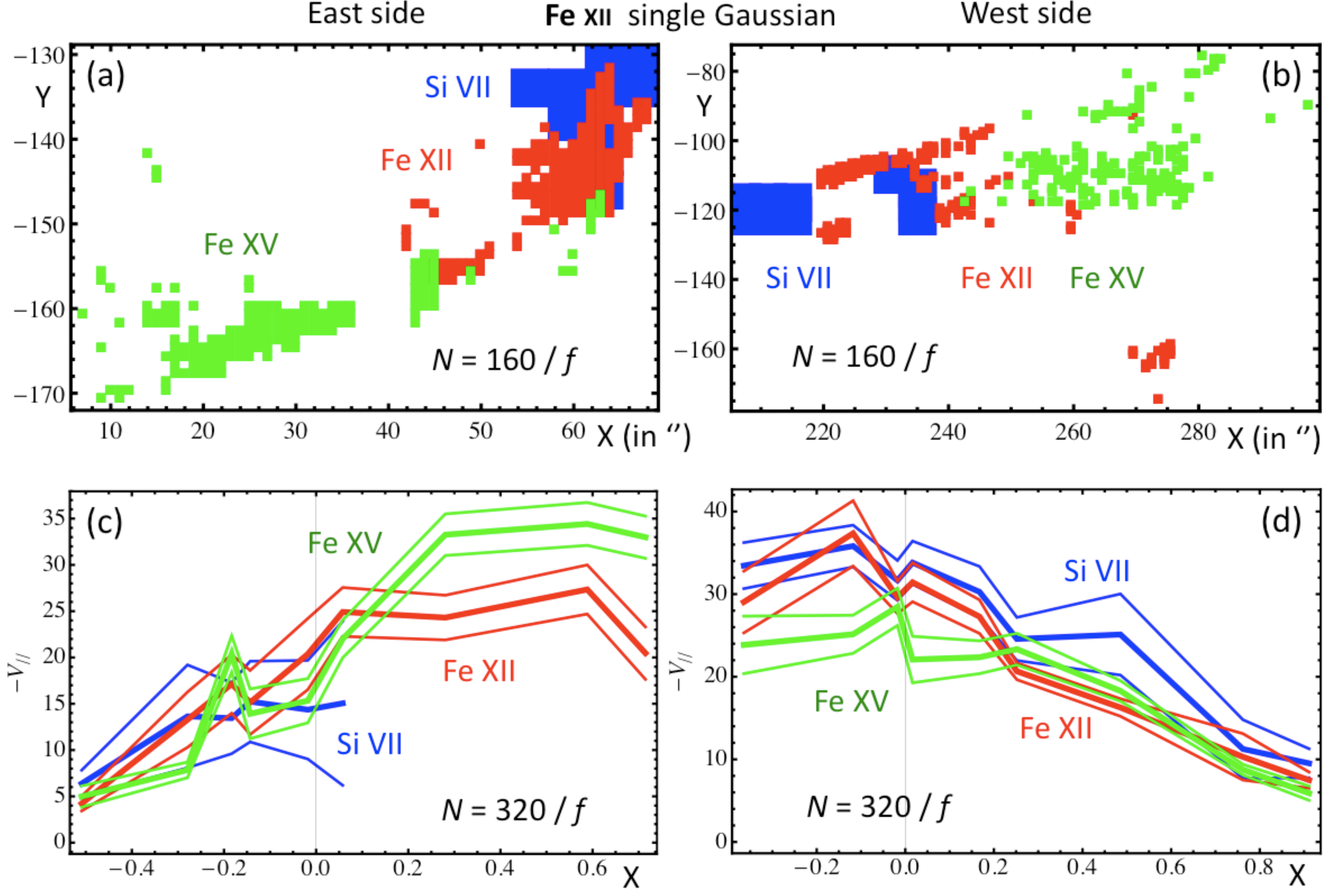}} 
\caption{
   {\bf (a,b)} Relative spatial positions of the fastest flows ($\Vo$) on 12 Dec. at 11:43 UT when the flows of both sides are viewed with a comparable perspective ($|\varphi + \delta|\approx 45 \degree$). The East-West position, $X$, is in seconds of arc.
   {\bf (c,d)} Apparent evolution of the mean $\Vo$ (thick curves) surrounded by their standard deviation (thin curves). $N$ is the number of data points retained with the highest upflows. $f=1$ is for original data (\FeXII\ and \FeXV ) and $f=16$ for rebinned data (\SiVII ). The East-West position, $X$, is normalized by the solar radius.
  }
 \label{fig:compare_Si_Fe}
\end{figure}  

\section{Spatial and Magnitude Spread of Velocities } \label{sec:Spread}

\subsection{Spatial Extent} \label{sec:Spread:Spatial}

A fan structure with upflows is observed on both sides of the AR in the \FeXII\ and \FeXV\ velocity maps (\fig{evol_Si_Fe}).  However, the fan structures in the \SiVII\ velocity map are globally more dominated by downflows.  This is especially true when the AR is far from the central meridian, say $|X|>300\arcsec$ (movies v\_compare* and v\_evolution*), but this is also partly true close to the central meridian (\fig{evol_Si_Fe}).  This is in agreement with previous results showing that the mean flow velocity is progressively changing from down- to upflows with increasing temperature in ARs (\eg\ \opencite{DelZanna09a}, \citeyear{DelZanna09b}; \opencite{Warren11}).  
Comparing with the results of the magnetic extrapolation, this implies that a large fraction of the \SiVII\ flows are on different loops than those observed in \FeXII\ and \FeXV\ spectral lines.   However, a part of the flows observed in \SiVII\ are upflows (\fig{evol_Si_Fe}).   Below we show that such flows are the counterpart of the upflows observed in the hotter Fe spectral lines.  Because of their lower heights and smaller extension (due to both a smaller plasma scale height and a decrease of the cross-section (convergence) of the flux tube at low height), such \SiVII\ flows can be easily mixed up with neighbouring downflows along the line of sight, reducing the observed upward velocity or even fully masking it.

  The \FeXII\ and \FeXV\ upflow velocities have a fan-like shape which is rooted in the \SiVII\ upflows (\fig{IvB_maps} and velocity movies).  The upflows are indeed structured along field lines within the AR (\fig{extrapol}) and connecting to the network (not shown). 
The spatial relationship between the highest upflows observed in the different wavelengths is summarized in \fig{compare_Si_Fe}a,b, at a time selected so that the viewing angles of flows on both AR sides are similar.   When the flows are directed toward the observer, the upflows at the three wavelengths appear closer together in the plane-of-sky (see movie v\_compare*).  Conversely, when the upflows are seen more from the side, the projected distance between the different wavelength increases, \ie\ the projection effect makes the temperature stratification more clear.  
However, as the observing direction is more orthogonal to the flows, the Doppler velocity decreases, and the blueshifted signal is getting more mixed with the back/foreground emissions.   This implies that the upflow locations become more uncertain as the upflows are observed more from the side. 
Therefore, we measure the distance between the mean position of the highest upflows observed at different wavelengths in an intermediate position ($|\varphi + \delta|\approx 45 \degree$, \fig{compare_Si_Fe}).  The \SiVII\ and \FeXII\ are separated along the flow direction by $\approx 15\arcsec$ on both AR sides, while the \FeXII\ and \FeXV\ are separated by $\approx 60\arcsec$ on the East side and only $\approx 30\arcsec$ on the West side. 

  The magnitude of the highest upflows in the different wavelengths is closely related (\fig{compare_Si_Fe}c,d).  When the upflows are pointing to the observer, they are slightly higher in \FeXV\ than in \FeXII\ on the East side while the reverse is present on the West side.  This could be only due to a different amount of bias present in the measured flows (such as too limited field of view and/or different contamination by the back/foreground emissions).  The same warning applies to \SiVII\ velocities.  Furthermore, on the East side the location of the highest upflows in \SiVII\ becomes a thin elongated region after the CMP (movie v\_evolution*). This thin region can be easily contaminated by surrounding redshifted flows, which are also brighter than the blueshifted flows.  Moreover, the weakness of the \SiVII\ line implies that we need a rebinning ($4\times 4$) in order to have sufficient signal over noise.  All of these factors contribute to weaker observed velocities in  \SiVII\ after CMP, rendering them unreliable (so they are not reported in \fig{compare_Si_Fe}c).   Finally, all in all we can at least conclude that the velocity magnitudes are comparable in all three spectral lines.  Before interpreting the differences, we need to see whether or not the results are robust, analysing other ARs.

\begin{figure} 
\centerline{\includegraphics[width=\textwidth]{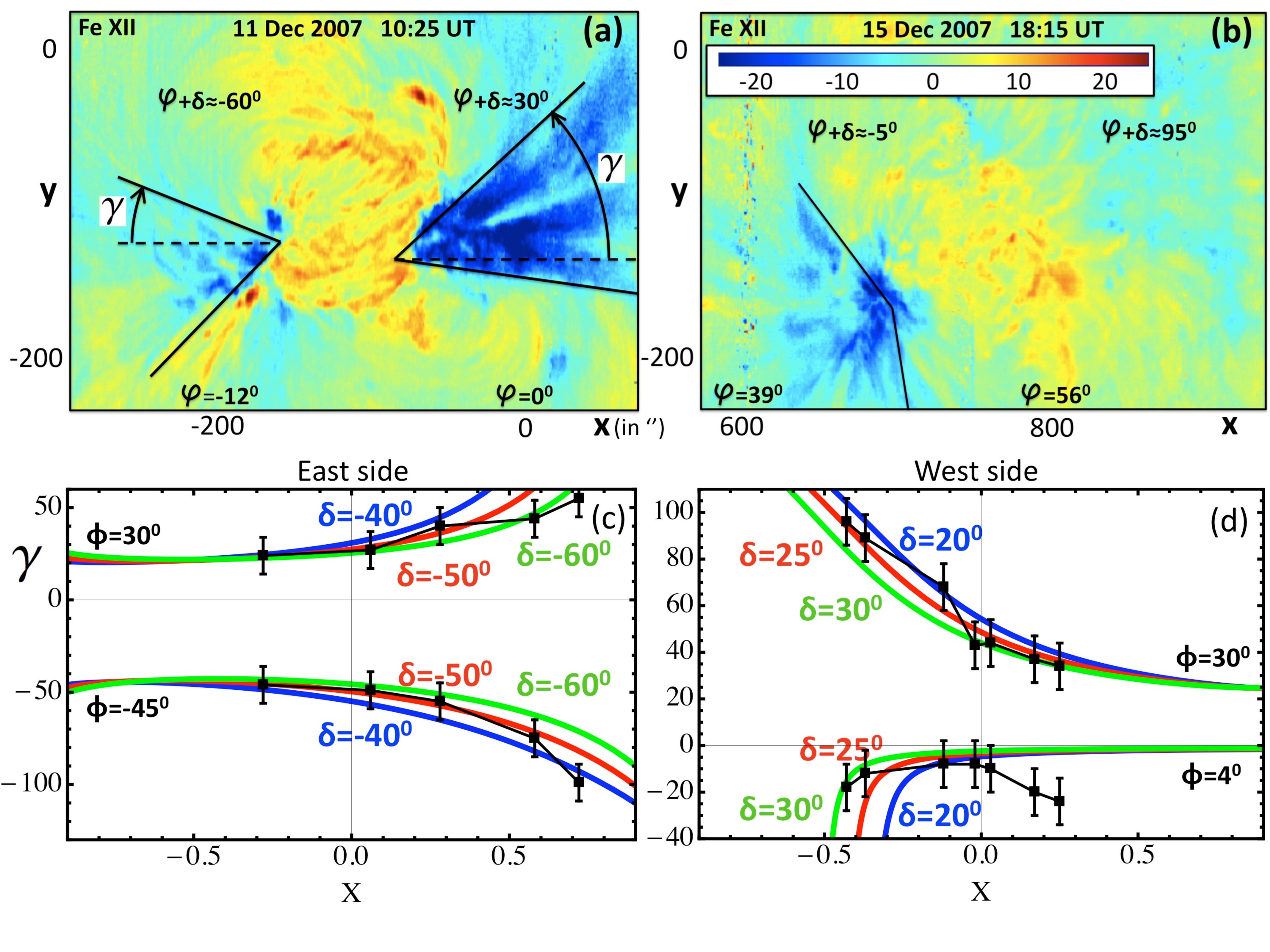}} 
     \vspace{-0.05\textwidth}   
\caption{{\bf (a,b)} Doppler velocity maps, computed from a single Gaussian fit to the \FeXII\ line, showing the angular extent of the flows on both sides of the AR.  In panel (a), the angle measuring the border direction of the flows, $\gamma$, is defined as positive on both sides. $\varphi+\delta$ is an estimation of the angle between the upflows and the LOS.
     {\bf (c,d)} Apparent evolution of the angular extent of the flows.  The expected evolution for steady flows, \eq{gamma}, is shown for three values of the East-West inclination $\delta$.  }
 \label{fig:angular_extent}
\end{figure}  

\subsection{North-South Angular Extent} \label{sec:Spread:NS}
  The observed angular extent of upflows is evolving with time, \eg\ the western upflow region is much broader earlier on (\textit{e.g.} \fig{IvB_maps}d) while the eastern upflow is much broader later on (\textit{e.g.} \fig{angular_extent}b, see also movie v\_evolution*). Indeed, such evolution is expected from a geometrical effect as the viewing angle of the fan is changing.  To quantify this, we measure the angle $\gamma$ of the fan at each side with respect to the East-West direction (\fig{angular_extent}), wherever possible.  It is a difficult task as the flows are fuzzy (no sharp boundaries), partly bent (along low intensity coronal loops), not always seen far enough (limited field of view, weak signal), and more importantly, the Doppler signal disappears as the flows are observed closer to their orthogonal direction (the best geometrical direction to derive the true angular width of the flows), see movie iv\_compare*.    

  Within the above limitations, we quantify the geometry of the flows using the coordinate systems introduced in \sect{Steady}, in particular in \eqs{Vdef}{Vlocal}.  
The tangent of the angle $\gamma$ (defining the observed flow borders) is: 
  \BE \label{eq:gamma} 
  \tan \gamma = \frac{\sqrt{1-Y^2} \; (Y \cos \delta \cos \phi + \sqrt{1-Y^2} \sin \phi)  }
                     {X \sqrt{1-Y^2} \cos \delta \cos \phi  
                     + \sqrt{1-X^2-Y^2} \sin \delta \cos \phi - X\;Y\;\sin \phi}  \, . 
  \EE
For low latitude, \textit{i.e.} small $Y$ value as for AR 10978, \eq{gamma} is simplified to:
  \BE \label{eq:gammaSimple} 
  \tan \gamma \approx \tan \phi / \sin (\varphi +\delta)   \, . 
  \EE
This implies that $|\gamma|=90 \degree$ when the angle between the flows and the LOS,
$\varphi +\delta$, vanishes, \ie\ any flow with a slight East-West component has an apparent North-South direction for this observing configuration.
        
 \fig{angular_extent}c,d shows that the measured $\gamma$ values on both fan sides and
both AR sides follow approximately \eq{gamma}.   On the eastern side, the inclination
is $\delta = -50 \pm 10 \degree$, while on the western side $\delta = 25 \pm 5 \degree$. Independently of the observed evolution of $V_{//}$ magnitude (\sect{Inclin}), this is a confirmation of the inclination away from the AR centre of the upflows, with a larger inclination in the following polarity.   Moreover, the $\delta$ values found are compatible with the values found in \sect{Inclin} (Table~\ref{inclinations}).

While we cannot observe the upflows close to their orthogonal orientation, we can still estimate their mean orientations and angular extents from the results shown in \fig{angular_extent}c,d.  On the eastern side  $\phi \approx  -7 \pm 32$ and $\phi \approx 17 \pm 13$ on the western side, so the fan on the eastern side is significantly broader than on the western side (by nearly a factor 3).  
Is it compatible with the magnetic extrapolation of \fig{extrapol}, and more precisely with the QSL extension?  In fact, the computation of the QSLs is not sufficient to determine where reconnection is happening along them. Indeed, thin current layers are built on these QSLs as the result of a temporal evolution, \eg, driven by boundary flows \cite{Aulanier05b}.  This implies that the reconnection location on the QSLs depends on their evolution.  Indeed, flare ribbons are typically present only along part of the thin QSLs (\eg, \opencite{Demoulin97}).  Then, we conclude that, without detailed information on the photospheric flows and a data-driven MHD simulation of the magnetic evolution, the QSL extension cannot be compared to the angular extent of the flows deduced above.  

On top of the above global estimation of the fan extent, some deviation from the theoretical curves are present. The largest one is for the southern boundary of the western fan (\fig{angular_extent}d).  This deviation is even larger for other sets of parameters ($\delta, \phi $), so this deviation cannot be interpreted with \eq{gamma}. We interpret it as a real temporal evolution of the flow pattern.

The above determination of the flow geometry allows us to estimate the contributions of $\Vt$ in $V_{//}$.  The relative contribution of $\Vt$ to $\Vr$ in \eq{V//} is $-\tan \theta \tan \phi / \cos \delta $.  AR 10978 is located at $Y\approx -100$\arcsec, so $\tan \theta \approx -0.1$. The flow structures are spread around the East-West direction (\figs{IvB_maps}{angular_extent}), so $\tan \phi$ has typically a low average value ($\leq 0.3$).  Finally, from the $\delta$ estimates (Table~\ref{inclinations}), $\cos \delta$ is larger than $0.6$.  Everything considered, this implies that the relative contribution of $\Vt$ to $\Vr$ is of the order of 5\%, so negligible.  Then, the least square fit of \eq{V//} to the data will provide $\Vr'\approx \Vr$ and $\Vp$, or equivalently $V$ and $\delta$ (with $\cos \phi \approx 1$). These values are provided at the top of the plots (Figures~\ref{fig:fit_FeXII_single}-\ref{fig:fit_FeXII_secondary}).  

\begin{figure}  
\centerline{\includegraphics[width=\textwidth]{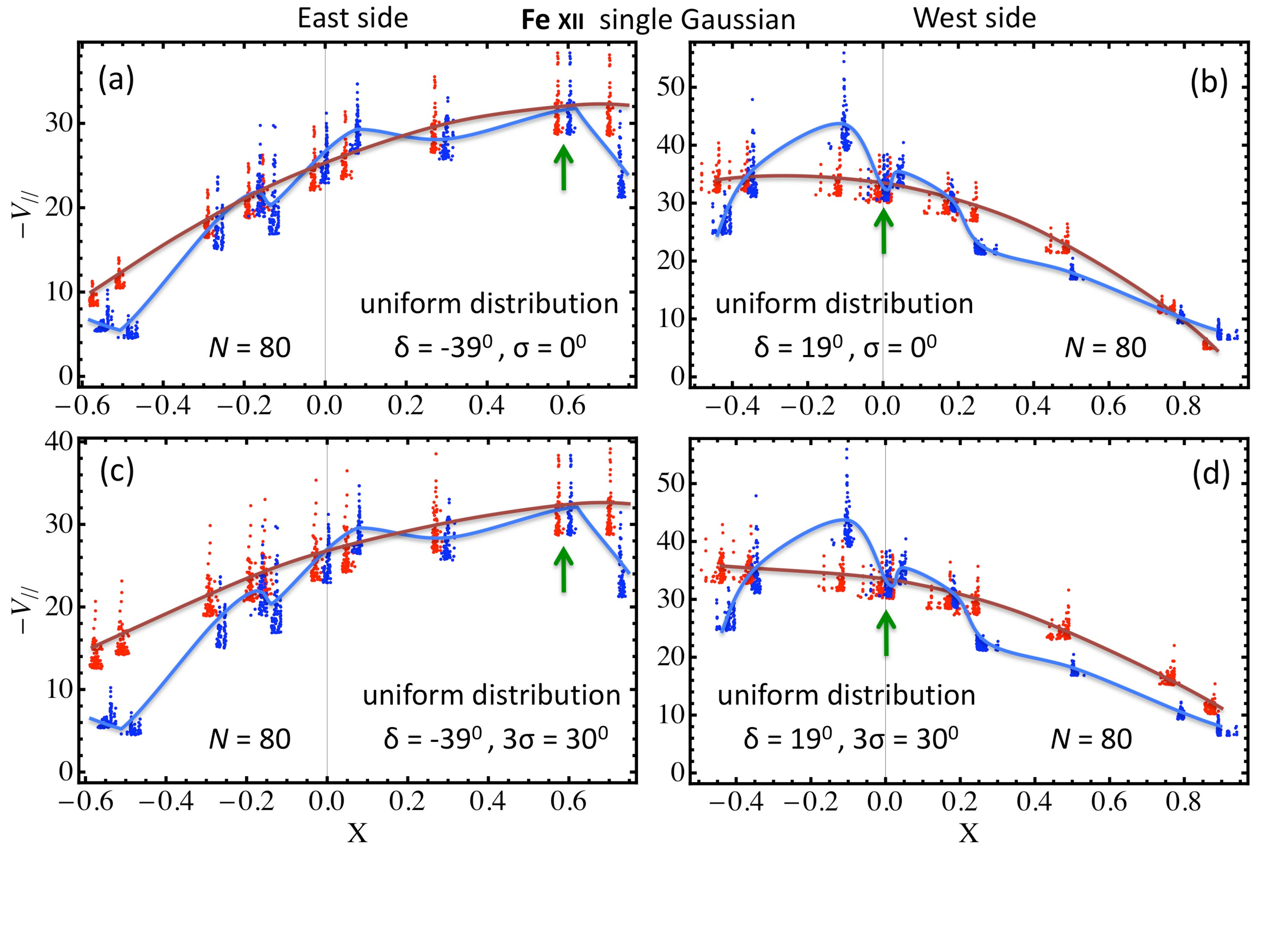}} 
     \vspace{-0.09\textwidth}   
\caption{Comparison of the observed (blue) and computed (red) $\Vo$ values as a function of $X$ (to avoid superposition a small $X$ shift is added/substracted, respectively). For each of the 10 data sets, the 80 highest $-\Vo$ values are retained.  For the model, $\Vo$ is computed from  the $V$ distribution derived from the data indicated by a green arrow and from a uniform distribution of flow directions (with a mean inclination of $\delta$ to the local vertical and with a standard deviation of $\sigma$). 
}
 \label{fig:mod_0_10}
\end{figure}  

\begin{figure}  
\centerline{\includegraphics[width=\textwidth]{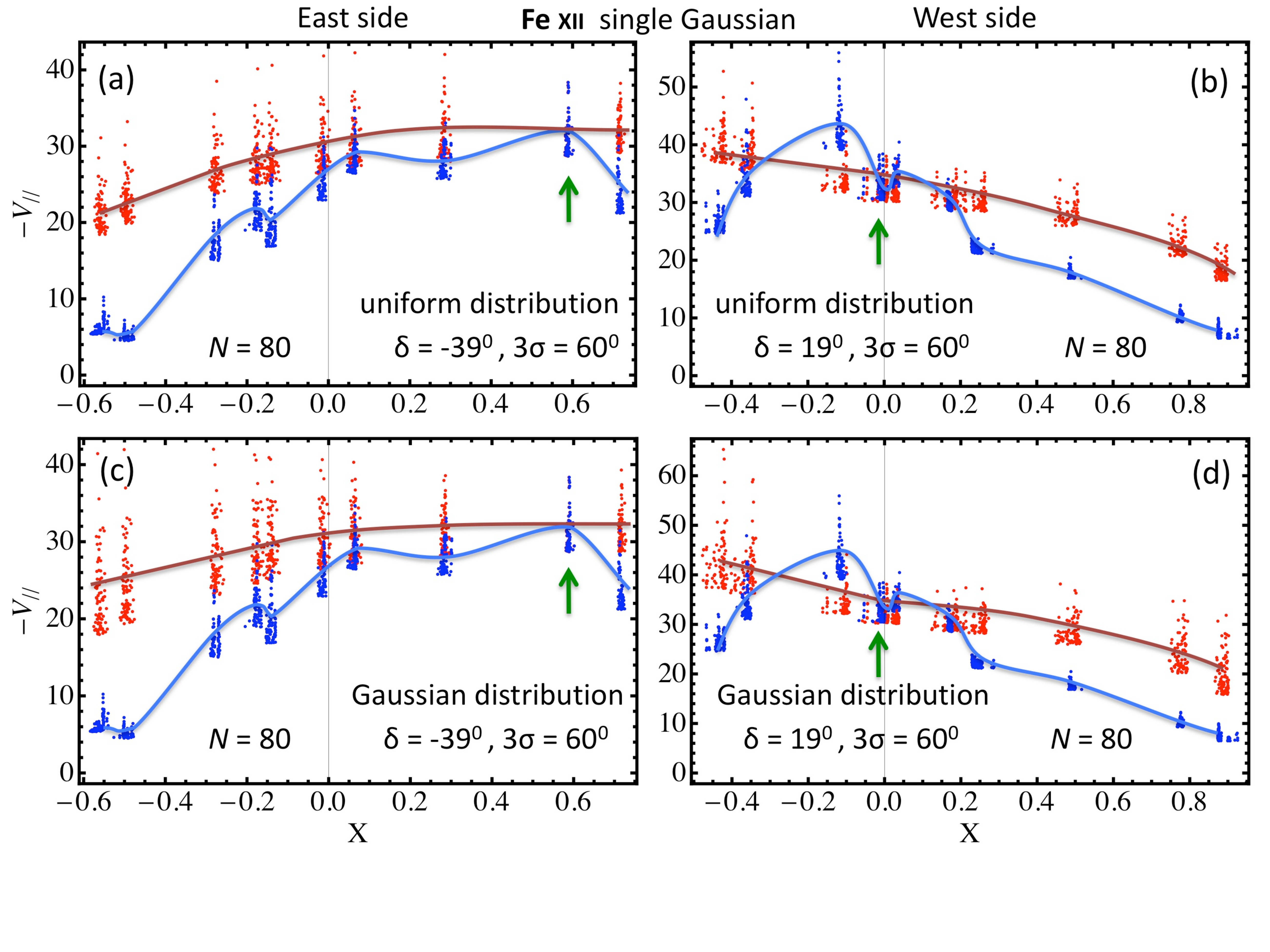}} 
     \vspace{-0.09\textwidth}   
\caption{  Comparison of the observed (blue) and computed (red) $\Vo$ values as a function of $X$ with the same drawing convention as in \fig{mod_0_10}.
Comparable results are obtained for a uniform and a Gaussian distribution of flow directions.  Such broad angular distributions are incompatible with observations. 
}
 \label{fig:mod_20}
\end{figure}  

\subsection{Vertical Angular Extent} \label{sec:Spread:Vertical}

   In \sect{Inclin} we assume that the flows have a common direction on each AR side.
However, the flow directions are expected to be spread around a mean direction.  While we estimate the North-South angular extent in the previous subsection, we cannot determine the vertical angular extent directly from the observations (as observations are done mostly from above the AR).  Still, we can do forward modelling, \ie\ we can suppose that a given angular distribution is present and compute its consequences on the LOS velocity.  More precisely, we investigate how different angular distributions of the velocities are affected by the solar rotation.  We found previously that the North-South angular extent of the flows is moderate (so that $\cos \phi \approx 1$). Moreover, the AR is close to the equator so that we can neglect the effect of the North-South angular extent of the flows on the longitudinal velocity $V_{//}$. Next, we suppose that the angular distribution, in the local vertical direction, is random and independent of the $V$ magnitude.  From these hypotheses we determine the expected distribution of the LOS velocity and compare it to the observed distribution.

The distribution of $V$ is best estimated from observations when the velocity is almost along the LOS, since then we minimize the effect of the angular distribution.  The results of \sect{Inclin:FeXII} determine which data set should be selected on each side (quoted below as the reference set, see the ones indicated by green arrows in \fig{mod_0_10}). On the West side, we avoid the data set showing the largest intrinsic evolution (located at $X\approx -0.1$), as it is not representative of the $V$ distribution of the other data sets. 
   
   It is worth noting that even if we use random distributions of inclinations, this does not imply that we suppose that inclinations are spatially randomly distributed within the observed flow structures, as follows.  Indeed, the $X,Y$ coordinates have a small spread within a data set (\eg\ \fig{fit_FeXII_single}), so that interchanging the assumed inclinations within the data points has a negligible effect on the distribution of $\Vo (X,Y)$.  Then, the observed flows could well have a coherent spatial organization, \ie\ with an inclination progressively changing with $X,Y$.  We are only modelling the distribution of inclinations independently of their spatial locations.   This is checked by using different random distributions (different seeds for the random generator) with the same $\delta$ and $\sigma$ values, so by redistributing the inclinations within the data set.   
   
As a first hypothesis, we select a uniform angular distribution with a mean inclination $\delta$, as determined in \sect{Inclin:FeXII} and a standard deviation $\sigma$.  From the observed velocities of the reference set and this angular distribution, we compute the expected distribution of $\Vo$.  For small values of $\sigma$ (say below $30\degree$), the angular distribution has a low effect on the derived $\Vo$.  The full angular extent of this uniform distribution is $\approx 3.4 \sigma$, so we quote below results with 3$\sigma$. 

On the AR's East side, the data are compatible with a negligible angular spread of the flows together with a small temporal evolution (\fig{mod_0_10}a).   An angular distribution spread with $3\sigma=30\degree$ shows a deviation from the observed velocities (\fig{mod_0_10}c).  This conclusion is confirmed when the mean and standard deviation of the distributions are compared. 
The same result is found for a broad range of the number $N$ of data points retained ($[20, 1000]$). Increasing $N$ has the same effect as in \sect{Inclin} (broader distributions and lower mean), still the model deviates from the observations in the same manner when $\sigma$ is increased.  The results on the West side are also similar; the observations are more compatible with a narrow angular distribution (\fig{mod_0_10}b,d) even though the more important flow evolution on this side weakens the result.

Next, we test if the results depend on the angular distribution by selecting a Gaussian distribution 
with the same mean inclination, $\delta$, and a standard deviation $\sigma$.  We find results very close to those with a uniform distribution.   Indeed small differences are visible only for large $\sigma$ values as the extended wings of the Gaussian distribution appear more clearly in the computed velocity.  Mainly a broader distribution of simulated velocities is obtained (\fig{mod_20}). 

For both angular distributions, the ones with a larger spread of flows, $3\sigma=60\degree$, are incompatible with the observations for both AR sides.  Indeed, broader distributions produce a weaker modulation than observed with solar rotation because we always select a fixed number of the largest LOS flows.  Finally, by selecting the $N$ largest velocities, we enhance the sensitivity of the selected velocities onto the vertical angular width of the fan compared to a simple projection effect on the full velocity distribution. 

\begin{figure} 
\centerline{\includegraphics[width=\textwidth]{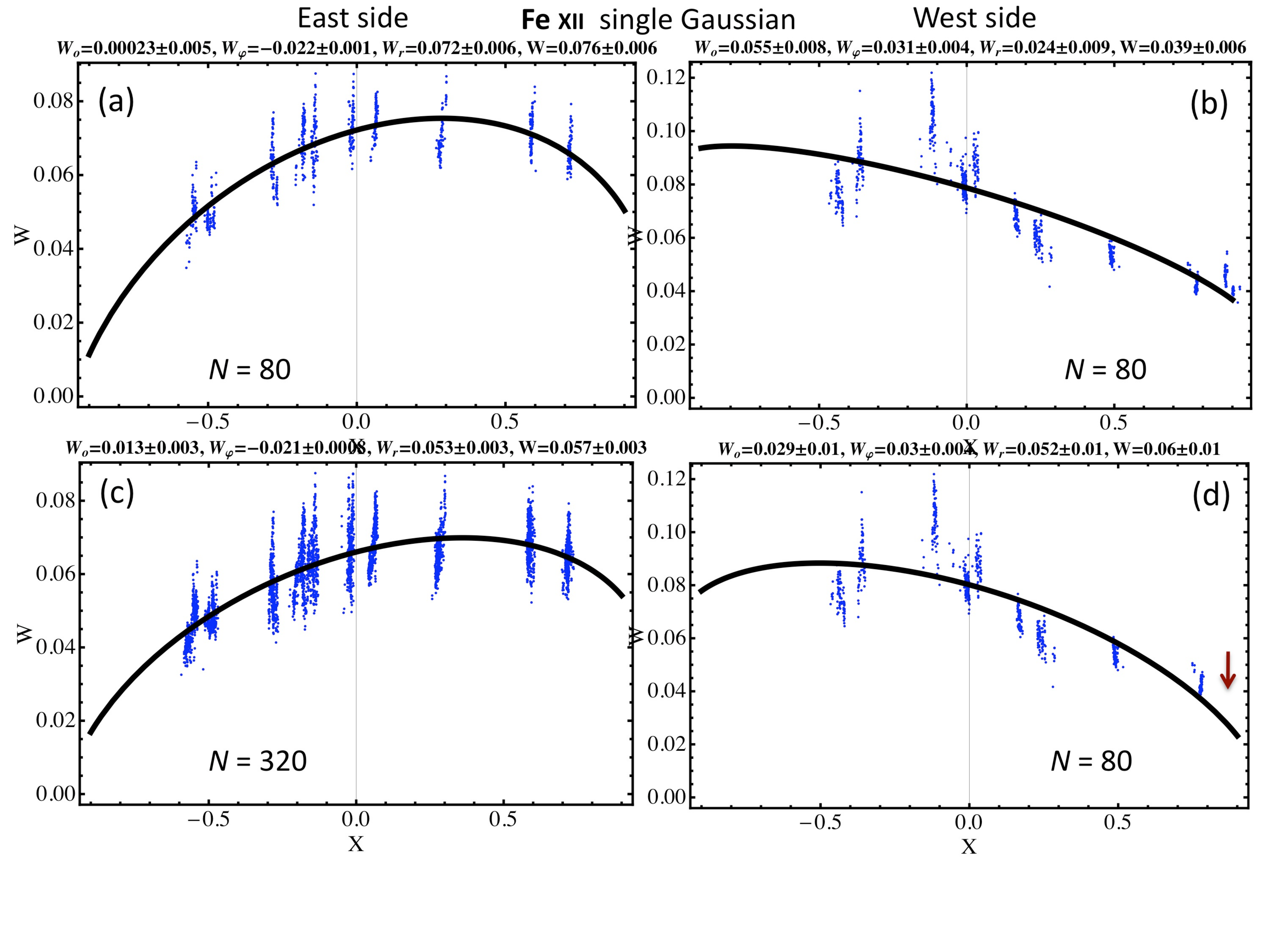}} 
     \vspace{-0.09\textwidth}   
\caption{Dependence of the line width, $W$ of \FeXIIA\ on the $X$ position (normalized to the solar radius) on the East (left panels) and West (right panels) sides of the AR.  The highest $-\Vo$ values are selected in each data set and the black line is the least square fit of \eq{W}.  The results of the fit are written on the top of each panel. In panel (d) the westernmost data set is removed (brown arrow). }
 \label{fig:fit_width}
\end{figure}  

\subsection{Velocity Magnitude Dispersion} \label{sec:Spread:Velocity}

The line width and the Doppler velocity of \FeXIIA\ are known to be correlated in AR upflows \cite{Doschek08,Hara08,Bryans10}.  Is this correlation due to the presence of directed flows with a dispersion of velocities, so that the spectral line is both blue-shifted and broadened by the same mechanism?  If it is so, then the line width should have a similar dependence to that of the mean velocity with respect to the $X$ coordinate of the flows.  In order to test this hypothesis, we perform a similar analysis as in \sect{Inclin} and we fit the data with
  \BE 
   W = W_{0} + \Wr' \sqrt{1-X^2-Y^2}  - \Wp \frac{X}{\sqrt{1-Y^2}}   \label{eq:W}
  \EE
where $W_{0}$, $\Wr'$ and $\Wp$ are the free coefficients of the fit. Compared to \eq{V//}, we added the constant $W_{0}$ to include all the width contributions which are independent of
position (like intrinsic and thermal broadening).  $\Wr'$ includes the radial and latidudinal broadening due to a dispersion of flow speed in these directions, while $\Wp$ takes into account the flow dispersion in the longitudinal direction. As with the LOS velocity, neglecting the latidudinal contribution, we can rewrite $\Wr$ and $\Wp$ as $W$ and $\delta_{\rm W}$ (like in \eq{Vlocal2}). $W$ is the line-width contribution due to the flow dispersion. However, since the broadening is due to a dispersion of velocities, we cannot interpret $\delta_{\rm W}$ as a measure of the flow inclination (as done in \sect{Inclin} for the LOS velocity).
Finally, as before, we select the $N$ largest upflows, so that the line width analysis is realized on the same data points as in \sect{Inclin} (for the same $N$ value). 

 On the eastern AR side, the variation of the line width is indeed well fitted with \eq{W} for a broad range of $N$ values (\fig{fit_width}a,c).  In contrast to the case of LOS velocity results (\fig{fit_FeXII_single}a), even the easternmost data follow well the global trend.  $W_{0}$ is found to be small, $\lesssim 0.01$ which is also of the order of the fit uncertainties, so it implies that $W_{0}$ is negligible and not measurable.  Then, we conclude that the line width is mainly due to line broadening by the dispersion of velocities from upward flows.  
 
 On the western AR side, the line width has also an important dependence on the $X$ coordinate (\fig{fit_width}b,d) as expected from \eq{W}.  However, the line width also shows an intrinsic temporal evolution (traced by the departure of the data from the fit) similar to the one found for the LOS velocity (\fig{fit_FeXII_single}b,d).  None of the datasets are particularly out of the least square fit, so the line width temporal evolution is less localized in time than that for the LOS velocities (\fig{fit_FeXII_single}b).  Removing the westernmost data set, as in \fig{fit_FeXII_single}d, implies a significant change in the fitted parameters, while the data still depart significantly from \eq{W}.  Then, on the West side, we cannot quantify $W_{0}$.  We can only conclude that a significant part of the line broadening is due to a dispersion of velocity from upward flows. 

The analysis of the line width shows that the line broadening is dominantly due to a large dispersion of the flow speeds.  Such dispersion ranges from negligible velocity up to more than 300 \kms\ (\sect{Inclin:Fast}).  Following the same logic as that presented in \sect{Spread:Vertical}, the vertical angular spread of the flows needs to be small in order to provide a large modulation of $W$ with $X$, as observed.  They also need to be nearly stationary to follow closely \eq{W} as obtained on the East side (\fig{fit_width}a,c).

\subsection{Origin of the upflows} \label{sec:Spread:Origin}

\begin{figure} 
\centerline{\includegraphics[width=\textwidth]{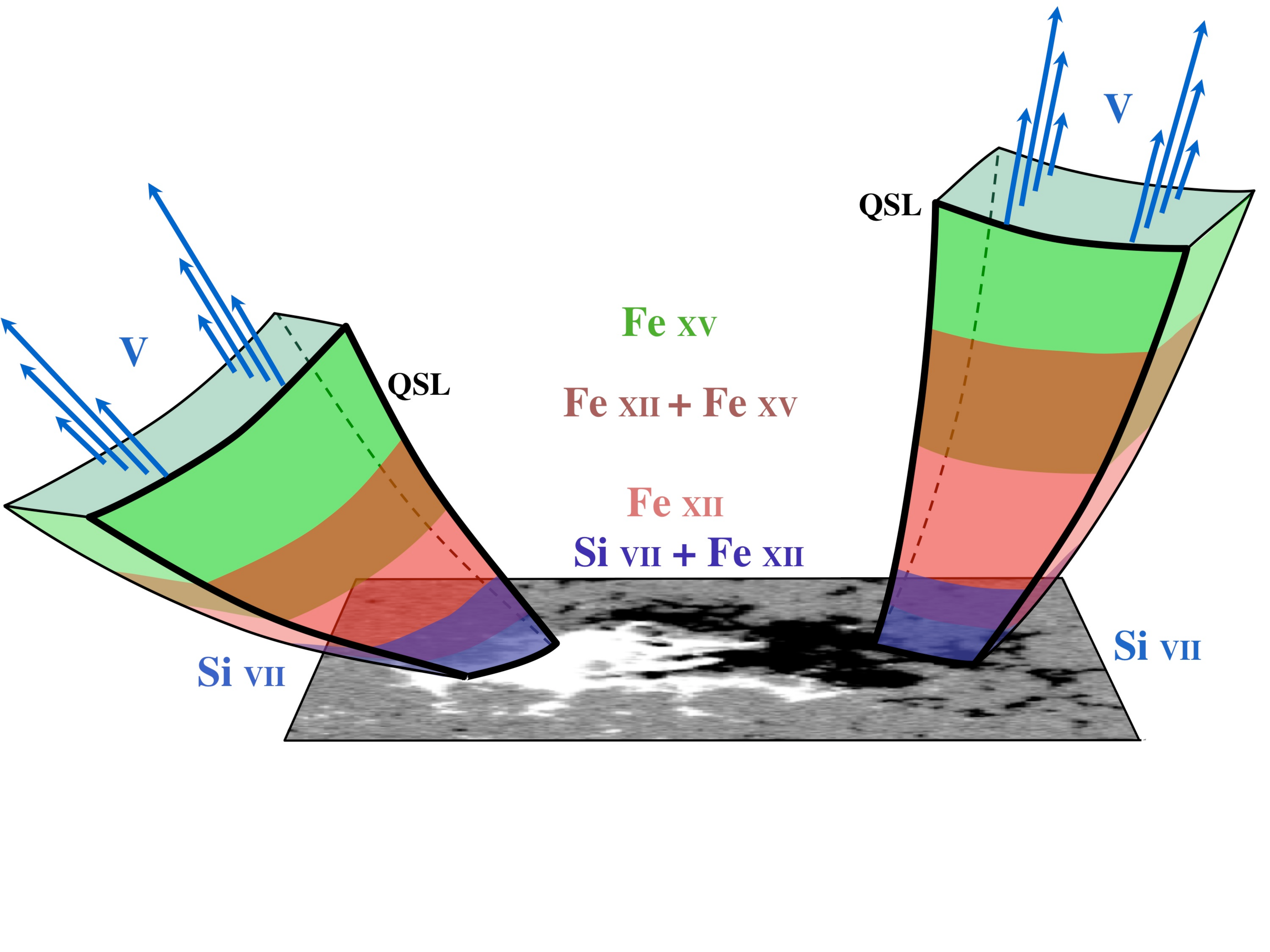}} 
     \vspace{-0.16\textwidth}   
\caption{Simplified schema of the upflows on top of a longitudinal magnetogram (taken on 12 Dec.). The AR is represented from a southern view point.  The upflows are formed due to reconnection at QSLs and as the plasma is being accelerated by a pressure gradient created by reconnecting loops of different plasma pressures, the temperature stratification of the solar atmosphere appears to show up in the blue-shifted flows observed at spectral lines of different temperatures.  Apart geometrical asymmetry (inclination and angular extension), the same physics is present in both AR polarities. }
 \label{fig:3DschemaFlows}
\end{figure}

We have shown in \sect{Inclin:Extrapol} that the upflows are located at the borders of computed QSLs, in agreement with previous studies \cite{Baker09a,vanDriel12}.  The flows have a fan-like structure with a relatively small angular width as sketched in \fig{3DschemaFlows}.  Moreover, the fastest upflows are supersonic and compatible with a pressure gradient due to overdense loops having reconnected with low density, larger loops as modeled by \inlinecite{Bradshaw11}.   All these results point towards a mechanism involving the reconnection of AR loops with surrounding larger loops as the driver of the observed upflows.   Such reconnection is occurring on the external part of  both AR polarities, where QSLs are present (and separatrices in the limit of infinitely thin QSLs).
While we observed only the consequence of reconnection (\ie, upflows), reconnection is expected to occur at QSLs as a result of the AR growth and dispersion.  Indeed, reconnection has been found at QSLs with evolving magnetic fields in numerical simulations (\eg, \opencite{Aulanier05b}; \opencite{Aulanier12}) and detected in laboratory experiments (\eg, \opencite{Gekelman12}). 

   On both sides of the AR, we also notice the tendency for the highest upflows observed in \FeXII\ and \FeXV\ to be closer to the AR centre, while weaker upflows are farther away (see movie v\_evolution*).  This could be an effect of a weaker Doppler signal as the flows become more dispersed away from the AR and more affected by background/foreground redshifted emissions.  This could also be the trace of the physics involved, as follows.  Just after reconnection, the plasma is rapidly accelerated by the upward pressure gradient.  Later, the over-pressure decreases as does the upflow speed (\fig{3DschemaFlows}).   So as the AR is becoming larger, reconnection is progressing toward the AR centre from both sides, and one expects the stronger flows closer to the AR centre, as observed.  Since, on both AR 10978 sides, we are seeing the upflows only from one side, a further analysis of other ARs is needed to check if the above expected asymmetry around the face-on view is well present or not.

\section{Conclusions}\label{sec:Conclusion}
 
 We select from the EIS dataset the AR with the largest temporal coverage in order to derive the physical properties of the upflows typically present on both sides of ARs. The Doppler velocity is generally obtained by a least square fit of a Gaussian function to the spectral profile (except when a double Gaussian fit is explicitly mentioned).   We find that the limb-to-limb evolution of the observed upflows is dominated by effects of the velocities projected onto the LOS, showing that the upflows have a strongly collimated stationary component.   

We analyse and model the highest upflows of each data sets.  The results are weakly dependent on the number $N$ of pixels retained.  From the least square fit of a stationary flow model to the data we deduce the East-West inclination, $\delta$, of the flow to the local vertical (\figs{fit_FeXII_single}{fit_SiVII_FeXV}).  We found coherent inclinations from spectral lines probing the corona between 0.6 and 2~MK.   These inclinations are also consistent with the mean inclinations of field lines rooted in QSLs related to the upflows (\fig{extrapol}).  The upflows are inclined away from the AR core on both magnetic polarities.
All results point to significantly more inclined flows in the following AR polarity ($\delta \approx -50\pm 10 \degree$) than on the leading polarity ($\delta \approx 20\pm 5 \degree$).
This is a consequence of the asymmetry in the magnetic polarities (leading/following).
The above results are schematically summarized in \fig{3DschemaFlows}.
 
The upflows observed in \SiVIIA , \FeXIIA , and \FeXVA\ are spatially related.  First, on both AR polarities, the flow locations partially overlap, with the \SiVII\ flows being closer to the AR core, the \FeXV\ flows being further from the AR core and the \FeXII\ flows in between (\fig{evol_Si_Fe}).  The spatial shift is present along the observed coronal loops as well as along the field lines computed from a magnetic field extrapolation.   The Doppler velocity has a similar temporal evolution in all spectral lines and the velocity magnitude is also comparable (\fig{compare_Si_Fe}, in contrast with previous studies (\eg\ see \opencite{DelZanna09a}, \citeyear{DelZanna09b}; \opencite{Warren11}) with a more global approach, whereas here we focus on the highest upflows).  All these results indicate that we observe practically the same upflows in different spectral lines, so different temperatures, at different locations along magnetic flux tubes, which is compatible with an idea of a stratified atmosphere set in motion (\fig{3DschemaFlows}).
 
Much faster upflows are detected with a double Gaussian fit of \FeXIIA\ (up to $\approx 300$ \kms ) than with a single Gaussian fit (up to $\approx 50$ \kms ).  These fast upflows are plausibly less inclined to the local vertical by a factor 2 on the AR East side (\fig{fit_FeXII_secondary}), while they have a comparable inclination on the West side compared to the flows detected with a single Gaussian fit.   However, the results of the double Gaussian fit can only be trusted in a very limited number of pixels where two main conditions are met. First, where upflow velocities are high enough to stand apart from the primary component, and second, where the emission measure is large enough to provide a reliable secondary component.    
 
The upflows have a fan-like structure in \FeXII\ and \FeXV .  The observed angular width of the fan is evolving with the AR position on the solar disc, becoming very broad when the flows are directed toward the observer (\fig{angular_extent}).   From this evolution and a stationary flow hypothesis, we deduce another estimation of the inclination of the flows to the local vertical which is compatible with the inclinations deduced from Doppler velocities and magnetic extrapolation.  We also deduce that the North-South angular extent of the flows is about $64 \degree$ and $26 \degree$ on the eastern and western AR sides, respectively.   Next, we use forward modelling to estimate the angular spread in the vertical direction. We infer that the observed velocities are only compatible with a narrow angular spread of the flows, \ie\ with a total angular extend below $30\degree$, both with a Gaussian and a uniform angular distribution (\figs{mod_0_10}{mod_20}).  This conclusion is stronger on the East side where the flows are more stationary. 

We further study the evolution of the line width with the strongest spectral line (\FeXIIA ) to have a large signal over noise ratio.  On the East side, as with the Doppler velocity, the observed line width of the highest flows is following well the expected evolution of a stationary flow only affected by the LOS projection (\fig{fit_width}).  The remnant line width, independent of the AR position, is very weak (within the error bars of the least square fit).  This implies that the line width is mostly due to a large dispersion of velocities in the main flow direction.   This explains the tied correlation found previously between Doppler velocity and line width \cite{Doschek08,Hara08,Bryans10}. On the West side, the results are compatible with the above conclusions, but they are less stringent because of an intrinsic temporal evolution of the upflows.

  From all the above results, the global picture is that the same upflows are detected in spectral lines around 1~MK (\fig{3DschemaFlows}).  They flow away from the AR core along field lines on both AR sides within a relatively narrow angular range but with a very broad velocity range. We see the same flows in spectral lines of different temperatures, but due to the thermal stratification of the atmosphere, their spatial locations and extents are different. Moreover, in the optically thin corona the contribution of Doppler backgrounds are significantly different for the \SiVII\ and Fe lines. As a result, the Doppler velocities and their spatial distribution appear different, while we are observing the same flows in all the lines.   While imaging with high temporal cadence has shown a short term variability (\textit{e.g.} \opencite{McIntosh09}; \opencite{Ugarte11}), these flows have also a strong stationary component, meaning that the driver is acting for several days, and probably weeks. The spatial locations of the flows within the magnetic configuration, as well as their small angular thickness (that we can only constrain from above), are compatible with a flow resulting from reconnection at QSLs, including separatrices \cite{Baker09a,vanDriel12}.  The highest flows are supersonic with a magnitude compatible with the simulation results of \inlinecite{Bradshaw11}.   We conclude that all the characteristics of the observed flows, including their 3D geometry explored in this article, are compatible with their generation by magnetic reconnection between dense AR loops having reconnected with long and low-density loops, then a pressure gradient drives supersonic flows which weaken in magnitude with time. This process is repeated with new loops creating a wide spread of velocities within the upflow regions. 

%
%
\begin{acks}
The authors thank the referee for helpful comments which improved the clarity of the paper. 
The research leading to these results has received funding from the European CommissionÕs Seventh Framework Programme under the grant agreement No. 284461 (eHEROES project). LvDGÕs work was supported by the Hungarian Research grant OTKA K-081421. CHM acknowledges financial support from the Argentinean grants PICT 2007-1790 (ANPCyT), UBACyT 20020100100733 and PIP 2009-100766 (CONICET). CHM is a member of the Carrera del Investigador Cient\'i fico (CONICET). PD and CHM thank ECOS-MINCyT for their cooperative science program A08U01. 
\end{acks}

%
 \bibliographystyle{spr-mp-sola-cnd}
 \bibliography{flows_AR.bib}  

\end{article} 
\end{document}